\documentstyle[sprocl]{article}
\input epsf.tex
\def\be{\begin{equation}}
\def\ee{\end{equation}}
\def\bea{\begin{eqnarray}}
\def\eea{\end{eqnarray}}

\def\ap{\alpha'}
\def\apm{\alpha'}
\def\Tr{{\rm Tr}}
\def\Z{{\bf Z}}
\def\R{{\bf R}}
\def\D{{\bf D}}
\def\rD{{\rm D}}
\def\rF{{\rm F}}
\newcommand{\zb}{{\bar{z}}}
\newcommand{\mbf}[1]{\mbox{\boldmath $#1$}}
\def\ha{{\textstyle{1\over 2}}}

\begin{document}

\title{TASI LECTURES ON D-BRANES}
\author{JOSEPH POLCHINSKI }
\address{Institute for Theoretical Physics\\
University of California, Santa Barbara, CA 93106-4030}

\maketitle\abstracts{
This is an introduction to the properties of D-branes, topological defects in
string theory on which string endpoints can live.  D-branes provide a simple
description of various nonperturbative objects required by string duality,
and give new insight into the quantum mechanics of black holes and the nature
of spacetime at the shortest distances.
The first two thirds of these lectures
closely follow the earlier ITP lectures hep-th/9602052, written with S.
Chaudhuri and C. Johnson.  The final third includes more extensive
applications to string duality.}
  
D-branes are extended objects, topological defects in a sense, defined by the
property that strings can end on them.  One way to see that these objects 
must be present in string theory is by studying the
$R \to 0$ limit of open string theory, where the different behaviors of open
and closed strings lead to a seeming paradox.\cite{dlp}  From the study of
this limit one can argue that the usual Type I, Type IIA and Type IIB string
theories are different states in a single theory, which also contains states
with arbitrary configurations of D-branes.  Moreover this argument is purely
perturbative, in that it does not require any strong-coupling continuation or
use of the BPS property.  

However, this subject remained obscure until the advent of string duality.  It
was then observed that D-branes have precisely the correct properties to fill
out duality multiplets,\cite{joeone} whose other members are fundamental string
states and ordinary field-theoretic solitons.  Further, they are much simpler
than ordinary solitons, and their bound states and other properties can be
understood more explicitly.  Even before the role of D-branes was
realized, the circumstantial evidence for string duality was 
extremely strong.  D-branes have extended this by providing a much more
complete and detailed dynamical picture.
Beyond the simple verification of weak/strong duality, D-branes have given
surprising and still developing insight into the quantum mechanics of black
holes and into the nature of spacetime at the shortest distance scales.

The first three lectures develop the basic properties of D-branes.  The
approach follows ref.~1, coming upon D-branes by way of $T$-duality. 
Lecture~1
is a review of open and unoriented bosonic strings, lecture~2 develops the
effect of $T$-duality in these theories, and lecture~3 extends these results
to the superstring.  The final two lectures develop various more advanced
ideas as well as applications to string duality.

These lectures are an expanded and updated version of an earlier set of
notes from lectures given at the ITP, which were written up
with Shyamoli Chaudhuri and Clifford Johnson.\cite{dnotes}  The first two
thirds of the present lectures closely
follow those notes.  The final third gives a more detailed treatment of the
application to string duality.  New or extended subjects include D-brane
actions, D-brane bound states, the connection to M theory and U-duality, and
D-branes as instantons, as probes, and as black holes.  Obviously, given
constraints of space and time, the discussion of each of these subjects is
somewhat abbreviated.

\section{Lecture 1: Open and Unoriented Bosonic Strings}

The weakly coupled $E_8 \times E_8$ heterotic string on a Calabi-Yau
manifold is strikingly similar to the grand unified Standard Model. 
Thus, the focus after the 1984 revolution was on this closed
oriented string theory.  We must therefore begin with an introduction to the
special features of open and unoriented strings.

\subsection{Open Strings}

To parameterize the open string world sheet, let the `spatial' coordinate
run $0 \leq \sigma^1 \leq \pi$. 
In conformal gauge, the action is
\be
{1\over4\pi\ap} \int_{\cal M} d^2\sigma
\, \partial_a X^\mu \partial_a X_\mu.
\ee
Varying $X^\mu$ and integrating by parts,
\be
\delta S = -{1\over2\pi\ap} \int_{\cal M} d^2\sigma \, \delta
X^\mu \partial^2
X_\mu\ +\ {1\over2\pi\ap} \int_{\partial M}d\sigma \, \delta
X^\mu \partial_n X_\mu \label{variation}
\ee
where $\partial_n$ is the derivative normal to the boundary.
The bulk term in the variation (\ref{variation}) gives
Laplace's equation (or the wave equation, if we have Minkowski signature on the
world-sheet.)
At the boundary, the only Poincar\'e invariant condition is the
Neumann condition
\be
\partial_n X^\mu=0. 
\ee  

The Dirichlet condition $X^\mu= {\rm constant}$
is also consistent with the equation of motion and we might study it
directly.  However, we have chosen to follow the approach of ref.~1, beginning
with the Neumann condition and finding that the Dirichlet condition is forced
on us by
$T$-duality.  This is a good pedagogic route, as the various properties of
D-branes make their appearance in a natural way.  It also shows that D-branes
are necessarily part of the spectrum of string theory, as one can start from
the ordinary open string vacuum and by a continuous change of parameters ($R
\to 0$) get to a state containing D-branes.

We should mention some early papers\,\cite{early} in which
Dirichlet conditions were investigated as a potentially interesting variation
of string theory in their own right.  The association of the Dirichlet
condition with an actual hyperplane in spacetime came later, in the context of
compactification of open string theory.\cite{latera}$^{\!,\,}$\cite{laterc}  It
was then observed that the hyperplane was a dynamical object\,\cite{dlp}, hence
`D-brane,' and that it could be produced via 
T-duality.\cite{dlp}$^{\!,\,}$\cite{hdual}$^{\!,\,}$\cite{gdual}

The papers in the previous paragraph kept the Neumann condition for the time
coordinate (and usually for some spatial coordinates) so that they defined
an object, the D-brane.  Fully Dirichlet conditions define the D-instanton. 
These boundary conditions were originally interpreted in terms of an off-shell
probe of string theory.\cite{offshell}  It was then proposed to modify string
theory by the inclusion of a gas of such objects, changing the short distance
structure.\cite{parton}  It was later argued\,\cite{joecomb} that such
D-instantons were a necessary part of string theory and carried the
combinatoric weight $e^{-O(1/g)}$ associated with large stringy
nonperturbative effects.\cite{shenk1}  It was also observed that the
D-instanton breaks half the supersymmetries.\cite{gdinst} 

The general solution to Laplace's equation with Neumann boundary conditions
is 
\be
X^\mu(z,\zb) = x^\mu - i\ap p^\mu
\ln(z\zb) + i\sqrt{\ap\over2} \sum_{m\neq0} {\alpha_m^\mu\over
m}(z^{-m}+\zb^{-m}),
\ee
where $x^\mu$ and $p^\mu$ are the position and momentum of the center of
mass.  As is conventional in conformal field theory, this has been
written in terms of the coordinate $z=e^{\sigma^2-i\sigma^1}$, so that
time runs radially.
The usual canonical quantization gives the commutators
\bea
[x^\mu,p^\nu] &=& i\eta^{\mu\nu}, \nonumber\\
{}[\alpha_m^\mu,\alpha_n^\nu] &=& m\delta_{m+n}\eta^{\mu\nu}.
\eea
The mass shell condition is
\be
M^2 = -p^\mu p_\mu = {1\over\ap}\left(\sum_{m=1}^\infty
mN_m-1\right).
\ee
Here $N_m$ is the number of excited oscillators in the
$m^{th}$ mode, and the $-1$ is the zero point energy of the physical
bosons.  Let us remind the reader of the mnemonic for zero point energies.  A
boson with periodic boundary conditions has zero point energy
$-{1\over24}$, and with antiperiodic boundary conditions it is $1\over48$.
For fermions, there is an extra minus sign.  For the bosonic string in 26
dimensions, there are 24 transverse (physical) degrees of freedom, for a
total of $-{24\over24} = -1$.

For example, the two lightest particle states and their vertex operators are
\bea
{\rm tachyon}&&
|{ k}\rangle, \quad M^2=-{1/\ap},\quad V=\exp({i{ k}\cdot { X}})
\nonumber\\
{\rm photon}&& \alpha^\mu_{-1}|{ k}\rangle, \quad
M^2 =0, \quad V^\mu=\partial_t X^\mu
\exp({i{ k}\cdot{ X}}). \label{particles}
\eea
Here $\partial_t$
is the derivative tangent to the string's world sheet boundary.  The
open string vertex operators are integrated only over the world-sheet
boundary as is evident in figure~1b.
\begin{figure}
\begin{center}
\leavevmode
\epsfbox{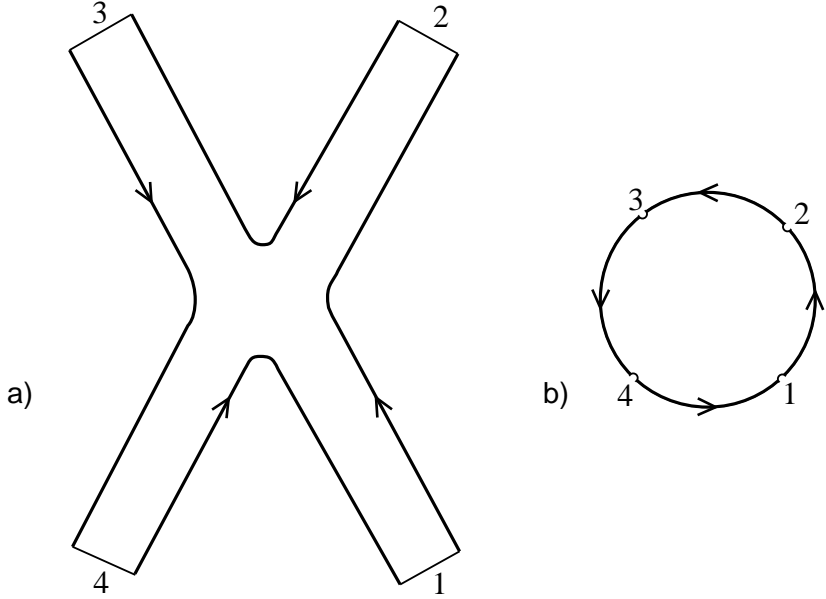}
\end{center}
\caption[]{a) Scattering of open strings.  b) Conformally transformed
world-sheet.}
\end{figure}

\subsection{Chan-Paton Factors}

It is consistent with spacetime Poincar\'e invariance and world-sheet
conformal invariance to add non-dynamical degrees of freedom to the ends of
the string. Their Hamiltonian vanishes so these {\it Chan-Paton} degrees of
freedom have no dynamics---an end of the string prepared in one of these
states will remain in that state.  So in addition to the usual Fock
space label for the state of the string,
each end is in a state $i$ or $j$ where the labels run from $1$ to $N$.
\begin{figure}
\begin{center}
\leavevmode
\epsfbox{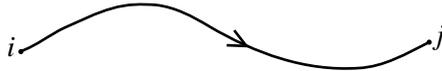}
\end{center}
\caption[]{Open string with Chan-Paton degrees of freedom.}
\end{figure}
The $n\times n$ matrices $\lambda^a_{ij}$ form a basis into which to
decompose a string wavefunction
\be
|{ k};a\rangle = \sum^N_{i,j=1}|{k},ij\rangle\lambda^a_{ij}.
\ee
These wavefunctions are the Chan-Paton
factors.\cite{chan}  Each vertex operator carries such a factor.
The tree
diagram for four oriented strings s shown in figure~1.
Since the Chan-Paton degrees of freedom are non-dynamical, the right end of
string 1 must be in the same state as the left end of string 2, and so on.
Summing over all basis states then gives a trace of the product of Chan-Paton
factors,
\be
\lambda^1_{ij}\lambda^2_{jk}\lambda^3_{kl}
\lambda^4_{li} = \Tr(\lambda^1\lambda^2\lambda^3\lambda^4). \label{factor}
\ee
Such a trace appears in each open string amplitude.
All such amplitudes are invariant under a the $U(N)$ symmetry\,\footnote
{The amplitudes are actually invariant under $GL(N)$, but this does not
leave the norms of states invariant.} 
\be
\lambda^i \ \to\ U \lambda^i U^{-1},
\ee
under which the endpoints transform as \mbf{N} and $\bar{\mbf{N}}$.
The massless vector vertex operator
$V^{a\mu}= \lambda^a_{ij}\partial_t X^\mu\exp(i{ k}\cdot { X})$
transforms as the adjoint under the $U(N)$ symmetry, so (as is generally the
case in string theory) this global symmetry of the world-sheet theory is
elevated to a gauge symmetry in spacetime.

The massless bosonic fields are the graviton
$G_{\mu\nu}$, dilaton $\phi$, antisymmetric tensor $B_{\mu\nu}$, and vector
$A^\mu$. The closed string coupling is related to the expectation value of the
dilaton field
$\phi_0$ by $g=e^{\phi_0}$.  In the low energy limit, the
massless fields satisfy equations of motion that may be obtained by
varying the action
\bea
S &=& \int
d^{10}x \left\{ {1\over2\kappa_0^2}
e^{-2\phi}\left(R+4(\nabla\phi)^2-{1\over12}
H_{\mu\nu\kappa}H^{\mu\nu\kappa}\right)\right. \nonumber\\
&&\qquad\qquad\qquad\qquad  -{c\over 4}e^{-\phi}\Tr
F_{\mu\nu}F^{\mu\nu} +O(\ap) \biggl\}\ . \label{stact}
\eea
This action arises at tree level in string perturbation theory.  
The dimensionful constant $\kappa_0$ is arbitrary, though in the context of
string duality there will be a natural additive normalization for $\phi$ and
so a natural value for $\kappa_0$.  The closed string kinetic terms are
accompanied by
$g^{-2}$ from the sphere and the open string kinetic terms by $g^{-1}$ from
the disk. The normalization of the open string action will be determined
later.

\subsection{Unoriented Strings}

Let us begin with the open string sector.  World sheet parity $\Omega$ takes
$\sigma^1 \to \pi-\sigma^1$ and acts on $z=e^{\sigma^2-i\sigma^1}$ as  
$z\leftrightarrow-\bar z$.  In terms of the mode expansion, $X^\mu(z,\bar z)
\to  X^\mu(-\bar z,- z)$ takes
\bea
x^\mu &\rightarrow& x^\mu \nonumber\\
p^\mu &\rightarrow& p^\mu \nonumber\\
\alpha^\mu_m &\rightarrow& (-1)^m\alpha^\mu_m\ .
\eea
This is a global symmetry of the open string theory above, but we can also
consider the theory that results when it is gauged.  When a discrete symmetry
is gauged, only invariant states are left in the spectrum.  The
familiar example of this is the orbifold construction, in which some
global world-sheet symmetry, usually a discrete symmetry of spacetime, is
gauged.
The open string tachyon is even and survives the 
$\Omega$-projection, while the photon does not, as
\bea
\Omega|{ k}\rangle &=& +|{ k}\rangle \nonumber\\
\Omega\alpha^\mu_{-1}|{ k}\rangle &=&  -\alpha^\mu_{-1}|{ k}\rangle.
\label{have}
\eea
We have made an assumption here about the overall sign of $\Omega$.  This
sign is fixed by the requirement that $\Omega$ be conserved in string
interactions, which is to say that it is a symmetry of the operator
product expansion (OPE).  The assignment~(\ref{have}) matches the symmetries
of the vertex operators~(\ref{particles}); in particular, the minus sign on
the photon is from the orientation reversal on the tangent derivative
$\partial_t$.

World-sheet parity reverses the Chan-Paton factors on the two ends of the
string, but more generally it may have some additional action on each
endpoint,
\be
\Omega\lambda_{ij}|{ k},ij \rangle\ \to\
\lambda^\prime_{ij} |{ k},ij \rangle, \quad
\lambda^\prime= M \lambda^{T} M^{-1}.
\ee
The form of the action on the Chan-Paton factor follows from the requirement
that this be a symmetry of general amplitudes such as~(\ref{factor}).

Acting twice with $\Omega$ squares to the identity on the fields, leaving
only the action on the Chan-Paton degrees of freedom. States are thus
invariant under 
\be
\lambda\to MM^{-T}\lambda M^TM^{-1}. \label{schur}
\ee  
The $\lambda$ must span a complete set of $N \times N$ matrices.  To see
this, observe that if strings $ik$ and $jl$ are in the spectrum for {\it any}
values of $k$ and
$l$, then so is the state $ij$.  This is because $jl$ implies $lj$ by CPT, and
a splitting-joining interaction in the middle gives $ik + lj \to ij + lk$.
But now eq.~(\ref{schur}) and Schur's lemma require $MM^{-T}$ to be
proportional to the identity,
so $M$ is either symmetric or antisymmetric.
This gives
two cases, up to choice of basis:\,\cite{sms}
\be
{\bf a.\ \ }{\rm Symmetric\colon}\ \ M=M^T=I_N
\qquad\qquad\qquad\qquad\qquad
\ee 
where $I_N$ is the $N\times N$ unit matrix.  In this case, for the photon
$\lambda_{ij}\alpha^\mu_{-1}|{ k}\rangle$ to be even under $\Omega$ and
therefore survive the projection, the Chan-Paton factor must be antisymmetric
to cancel the transformation of the oscillator state.  So
$\lambda=-\lambda^T$, giving the gauge group is $SO(N)$.
\be
{\bf b.\ \ }{\rm Antisymmetric\colon}
\ \ M=-M^T=i\left[ \begin{array}{cc}
0&I_{N/2}\\ -I_{N/2}&0 \end{array}\right].
\ee
In this case, $\lambda=-M\lambda^TM$, which defines the gauge
group $USp(N)$.\footnote{In the notation where $USp(2)\equiv SU(2)$.  This
group is also denoted $Sp(N/2)$.}

Now consider the closed string sector. For closed strings, we have the 
familiar mode expansion 
$X^\mu(z,\zb)=X^\mu(z)+X^\mu(\zb)$ with
\bea
X^\mu(z)&=&x^\mu+i\sqrt{\ap\over2}
\left(-\alpha_0^\mu
\ln z+\sum_{m\neq0}{\alpha^\mu_m\over mz^m}\right), \nonumber\\
X^\mu(\zb)&=&{\tilde x}^\mu+i\sqrt{\ap\over2}\left(-{\tilde\alpha}_0^\mu
\ln \zb+\sum_{m\neq0}{{\tilde\alpha}^\mu_m\over m\zb^m}\right).
\label{cmodes}
\eea
The theory is invariant under a  world-sheet parity symmetry
$\sigma^1\to-\sigma^1$.
For a closed string, the action of $\Omega$ is thus to reverse the right- and
left-moving oscillators,
\be
\Omega\colon\quad\alpha^\mu_m\leftrightarrow{\tilde\alpha}^\mu_m.
\ee
For convenience, parity is here taken to be $z \to \bar z$, differing by a
$\sigma^1$-translation from $z \to -\bar z$.  This is a global symmetry,
but again we can gauge it.  We have
${\Omega|{ k}\rangle=|{ k}\rangle}$, and so the tachyon remains in the
spectrum.  However
\be
\Omega\alpha^\mu_{-1}{\tilde\alpha}^\nu_{-1}|{
k}\rangle= {\tilde\alpha}^\mu_{-1}\alpha^\nu_{-1}|{ k}\rangle,
\ee
so only states symmetric under $\mu\leftrightarrow\nu$ survive from this
multiplet, i.e. the graviton and dilaton.  The antisymmetric tensor is
projected out.

One can also think about the gauging of $\Omega$ in terms of world-sheet
topologies.  When a world-sheet symmetry is gauged, 
a string carried around a closed curve on the world-sheet need only come
back to itself up to a gauge transformation.  Gauging world-sheet parity
thus implies the inclusion of unoriented world-sheets.
Figure~3 shows an example.  The oriented one-loop closed string amplitude
comes only from the torus, while insertion of the projector
${1\over2}\Tr(1+\Omega)$ into a closed string one-loop
amplitude will give the amplitudes on the torus plus Klein bottle.
\begin{figure}
\begin{center}
\leavevmode
\epsfbox{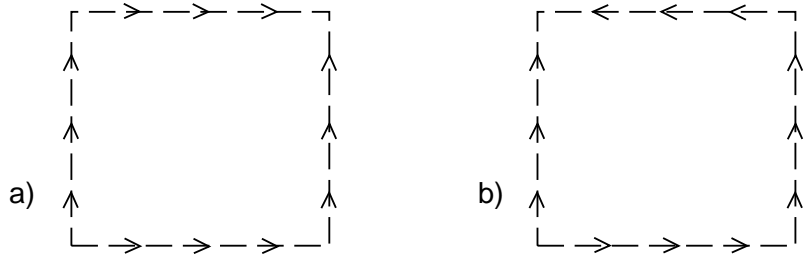}
\end{center}
\caption[]{a) Torus formed by identifying opposite edges.  b) Klein bottle
formed by identification with a twist.}
\end{figure}
Similarly, the unoriented one-loop open string amplitude comes from the 
annulus and M\"obius strip.  We will discuss these amplitudes in
more detail later.

The lowest order unoriented amplitude is the projective
plane, which is a disk with opposite points identified.
\begin{figure}
\begin{center}
\leavevmode
\epsfbox{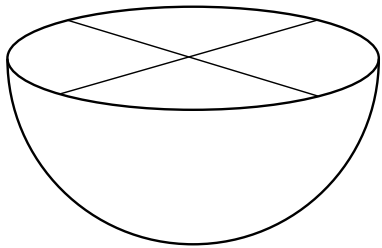}
\end{center}
\caption[]{Projective plane formed by identifying opposite points on the
disk as shown.  This can also be regarded as a sphere with a crosscap
inserted.}
\end{figure}
The
projective plane can be thought of as a sphere with a crosscap inserted,
where a crosscap is a
circular hole with opposite points identified.  A sphere with
two crosscaps is the same as a Klein bottle; this representation will be
useful and will be explained further in section~2.6.

Gauging world-sheet parity is similar to the usual orbifold construction,
gauging an internal symmetry of the world-sheet theory.  One difference is
that there is no direct analog of the twisted states because the Klein
bottle does not have the modular transformation $\tau \to -1/\tau$. 
On the torus, the projection operator implies the insertion of twists in the
timelike (vertical) direction of figure~3a; rotating the figure by
$90^\circ$, these become twists in the spacelike direction, implying twisted
states in the spectrum.  But if the Klein bottle of figure~3b is rotated by
$90^\circ$, the directions of time on the two opposite edges do not match and
there is no interpretation in terms of intermediate states in this channel.

Various authors have identified open strings as the twisted states
of the $\Omega$ projection.\cite{opentwist,laterc}  There are some definite
senses in which these are parallel, but the analogy is not exact and we find it
dangerous in that it might lead one to be less than general.

\section{Lecture 2: $T$-Duality}

\subsection{Self-Duality of Closed Strings}

For closed strings, let us first study the zero modes.  The mode
expansion~(\ref{cmodes}) is
\be
X^\mu(z,\zb) = x^\mu + \tilde x^\mu -i\sqrt{\ap\over2}
(\alpha^\mu_0+{\tilde\alpha}^\mu_0)\sigma^2 + \sqrt{\ap\over2}
(\alpha^\mu_0-{\tilde\alpha}^\mu_0)\sigma^1
+ {\rm oscillators}.
\ee
Noether's theorem gives the spacetime momentum of a string as
\be
p^\mu =
{1\over{\sqrt{2\ap}}}(\alpha^\mu_0 + {\tilde\alpha}^\mu_0).
\ee
Under $\sigma^1 \to \sigma^1+2\pi$,
the oscillator term are periodic and $X^\mu(z,\zb)$
changes by $2\pi\sqrt{(\ap/2)}(\alpha^\mu_0-{\tilde\alpha}^\mu_0).$
For a non-compact spatial direction $\mu$, $X^\mu(z,\zb)$ is single-valued,
and so 
\be
\alpha^\mu_0={\tilde\alpha}^\mu_0=\sqrt{\ap\over2}p^\mu. \label{plepr}
\ee
Since vertex operators must leave the space~(\ref{plepr}) invariant,
they may contain only the combination $x^\mu + \tilde x^\mu$.

For a compact direction of radius $R$, say $\mu=25$, $X^{25}$ has period
$2\pi R$. The momentum $p^{25}$ can take the values $n/R$.
Also, under $\sigma^1\sim\sigma^1+2\pi$, $X^{25}(z,\zb)$
can change by $2\pi wR$.  Thus
\bea
\alpha^{25}_0+{\tilde\alpha}^{25}_0 &=&
{2n\over R}\sqrt{\ap\over2} \nonumber\\
\alpha^{25}_0-{\tilde\alpha}^{25}_0 &=& 
\sqrt{2 \over \ap}wR
\eea
and so
\bea
\alpha^{25}_0 &=& \left({n\over R}+{wR\over\ap}\right)
\sqrt{\ap\over2} \nonumber\\
{\tilde\alpha}^{25}_0 &=&
\left({n\over R}-{wR\over\ap}\right)\sqrt{\ap\over2}.
\eea

Turning to the mass spectrum, we have
\bea
M^2\ =\ -p^\mu p_\mu &=&
{2\over\ap}(\alpha_0^{25})^2+{4\over\ap}
({ N}-1) \nonumber\\
&=& {2\over\ap}({\tilde\alpha}_0^{25})^2+{4\over\ap}
( {\bar N}-1).
\eea
Here $\mu$ runs only over the non-compact dimensions, $N$ is the
total level of the left-moving excitations, and $ \bar N$ the total
level of the right-moving excitations.  As $R \to \infty$, all states with
$w\neq 0$ become infinitely massive, while the $w=0$ states with all
values of $n$ go over to a continuum.  As $R\to 0$, all states with $n
\neq 0$ become infinitely massive.  In field theory this is all that would
happen---the surviving fields would be independent of the compact
coordinate, so the effective dimension is reduced.  In closed string theory
things work quite differently: the $n=0$ states with all $w$ values form a
continuum as $R\to 0$, because it is very cheap to wind around the small
circle.  In the $R\to 0$ limit, the compactified dimension reappears!
This is the first and still most striking indication that strings see
spacetime geometry differently from the way we are used to.  Indeed, many
other examples of `stringy geometry' or `quantum geometry' are closely
related to this.\cite{mirror}

The mass spectra of the theories at radius $R$ and $\ap/ R$ are identical
when the winding and Kaluza-Klein modes are interchanged
$n \leftrightarrow w$,\cite{tdual}
which takes
\be
\alpha^{25}_0 \rightarrow\alpha^{25}_0, \quad
{\tilde\alpha}_0^{25} \rightarrow -{\tilde\alpha}_0^{25} 
\label{tzemo}.
\ee
The interactions are identical as well.\cite{nairet}
Write the radius-$R$
theory in terms of 
\be
X^{\prime25}(z,\zb)=X^{25}(z)-X^{25}(\zb)\ . \label{onesidep}
\ee
The energy-momentum tensor and OPE and therefore all of the correlation
functions are invariant under this rewriting.  The only change, evident
from eq.~(\ref{tzemo}), is that the zero mode spectrum in the new variable
is that of the $\ap/R$ theory.  In other words, these theories are
physically identical; 
$T$-duality, relating the $R$ and $\ap/R$ theories, is an exact
symmetry of perturbative closed string theory.  Note that the
transformation~(\ref{onesidep}) can be regarded as a spacetime parity
transformation acting only on the right-moving degrees of freedom.

This duality transformation is in fact an {\it exact} symmetry of closed 
string theory.\cite{dhs}  To see why, recall from Ooguri's lectures
the appearance of an $SU(2)_L \times SU(2)_R$ extended gauge symmetry 
at the self-dual radius.  Additional left- and right-moving 
currents are present at this radius in the massless
spectrum, 
\bea
SU(2)_L\colon&&\partial X^{25}(z),\ \ \exp(\pm 2iX^{25}(z)/\sqrt{\ap})
\nonumber\\
SU(2)_R\colon&&\bar\partial X^{25}(z),\ \ \exp(\pm 2iX^{25}(\zb)/\sqrt{\ap})\
.
\eea
The marginal operator for the change of radius, $\partial X^{25} 
{\bar\partial} X^{25}$, transforms as a $({\bf 3},{\bf 3})$ and so a
rotation by $\pi$ in one of the $SU(2)$'s transforms it into minus itself.
The
transformation $R \to \ap/R$ is therefore a ${\bf Z}_2$ subgroup of
the $SU(2) \times SU(2)$.  We may not know what non-perturbative string
theory is, but it is a fairly safe bet that it does not violate spacetime
gauge symmetries explicitly, else the theory could not be
consistent.\footnote{Note that the ${\bf Z}_2$ is already spontaneously
broken, away from the self-dual radius.}  This is a simple example of an
idea which plays a prominent role in the study of string duality: that
arguments based on consistency in the low energy field theory place strong
constraints on the non-perturbative behavior of strings.

It is important to note that $T$-duality acts nontrivially on the
dilaton.\cite{gvb}  By the usual dimensional reduction, the effective
25-dimensional coupling is
$e^{\phi} R^{-1/2}$.  Duality requires this to be equal to 
$e^{\phi'} R'^{-1/2}$, hence
\be
e^{\phi'} = e^{\phi} R^{-1} {\ap}^{1/2}. \label{tdil}
\ee

\subsection{$T$-Duality of Open Strings}

Now consider the $R \to 0$ limit of the open string spectrum.  Open
strings cannot wind around the periodic dimension; they have no quantum
number comparable to $w$.  So when $R \to 0$ the states with nonzero
momentum go to infinite mass, but there is no new continuum of states.
The behavior is as in field theory: the compactified dimension disappears,
leaving a theory in $D-1$ spacetime dimensions.  The seeming paradox
arises when one remembers that theories with open strings always have
closed strings as well, so that in the $R\to 0$ limit the closed strings
live in $D$ spacetime dimensions but the open strings only in $D-1$.

One can reason out what is happening as follows. The string in the interior
of the open string is the same `stuff' as the closed string is made of,
and so should still be vibrating in $D$ dimensions.  What distinguishes
the open string is its endpoints, and these are restricted to a $D-1$
dimensional hyperplane.  Indeed, this follows from the duality
transformation~(\ref{onesidep}).  The Neumann condition $\partial_n
X^{25}$ for the original coordinate becomes $\partial_t X'^{25}$ for the
dual coordinate.\cite{dlp}$^{\!-\,}$\cite{gdual}  This is the Dirichlet
condition: the
$X^{25}$ coordinate of the endpoint is fixed, so the endpoint is constrained
to lie on a hyperplane.

In fact, all endpoints are constrained to lie on the same hyperplane.  To
see this, integrate 
\bea
X'^{25}(\pi) - X'^{25}(0) &=& \int_0^\pi d\sigma \partial_\sigma X'^{25}
\ =\ i \int_0^\pi d\sigma \partial_\tau X^{25} \nonumber\\
&=& 2\pi \ap p^{25}\ =\ \frac{2\pi \ap n}{R}\ =\ 2\pi n R'. \label{deltax}
\eea
That is, $X'^{25}$ at the two ends is equal up to an integral multiple
of the periodicity of the dual dimension, corresponding to a string that
winds as in figure~5.
\begin{figure}
\begin{center}
\leavevmode
\epsfbox{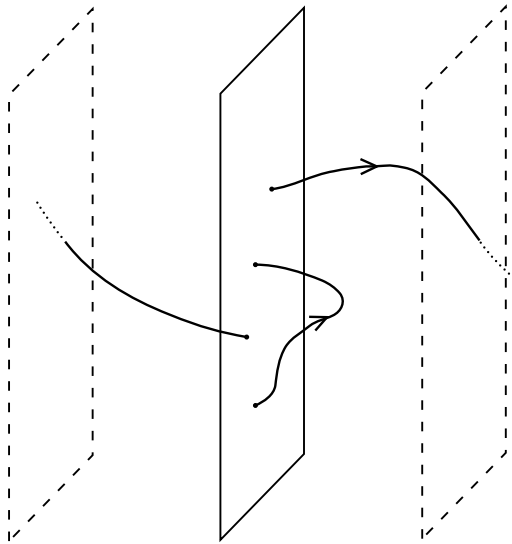}
\end{center}
\caption[]{Open strings with endpoints attached to a hyperplane.  The
dashed planes are periodically identified.  The strings shown have
winding numbers zero and one.}
\end{figure}
For two different open strings, consider the connected world-sheet that
results from graviton exchange between them.  One can carry out the same
argument~(\ref{deltax}) on a path connecting any two endpoints, so all
endpoints lie on the same hyperplane.  The ends are still free to move in
the other $D-2 = 24$ spatial dimensions.

\subsection{$T$-Duality with Chan-Paton Factors and Wilson Lines}

Now we study the effect of
Chan-Paton factors.\cite{joecomb}
Consider the case of $U(N)$, the oriented open string. In compactifying the
$X^{25}$ direction, we can include a Wilson line $A_{25}={\rm
diag}\{\theta_1,\theta_2,\ldots,\theta_N\}/2\pi R$, 
generically breaking $U(N) \to U(1)^N$.  Locally this is pure gauge,
\be
A_{25} = -i\Lambda^{-1}\partial_{25}\Lambda,\qquad
\Lambda={\rm diag}\{e^{ i X^{25}\theta_1/2\pi R},e^{ i X^{25}\theta_2/2\pi
R}, \ldots , e^{ i X^{25}\theta_1/2\pi R} \}\ .
\ee
One can then gauge $A_{25}$
away, but the gauge transformation is not periodic and the fields now pick up
a phase
\be
{\rm diag}\left\{ e^{-i\theta_1}, e^{-i\theta_2}, \ldots, e^{-i\theta_N}
\right\} \label{wilphase}
\ee
under $X^{25} \to X^{25} + 2\pi R$.  What is the effect in the dual theory?
Due to the phase~(\ref{wilphase}) the open string momenta now in general have
fractional parts.  Since the momentum is dual to the winding number, we
expect the fields in the dual description to have fractional winding
number, meaning that their endpoints are no longer on the same hyperplane.  Indeed,
a string whose endpoints are in the state $|ij\rangle$ picks up a phase
$e^{i(\theta_j - \theta_i)}$, so the momentum is $(2\pi n + \theta_j -
\theta_i)/2\pi R$.  The calculation~(\ref{deltax}) then gives
\be
X'^{25}(\pi) - X'^{25}(0) = (2\pi n + \theta_j - \theta_i) R'.
\ee
In other words, up to an arbitrary additive normalization, the endpoint in
state $i$ is at 
\be
X'^{25}\ =\ \theta_i R' \ =\ 2\pi\ap A_{25,ii}. \label{dualpos}
\ee  
There are in general $N$
hyperplanes at different positions as depicted in figure~6.
\begin{figure}
\begin{center}
\leavevmode
\epsfbox{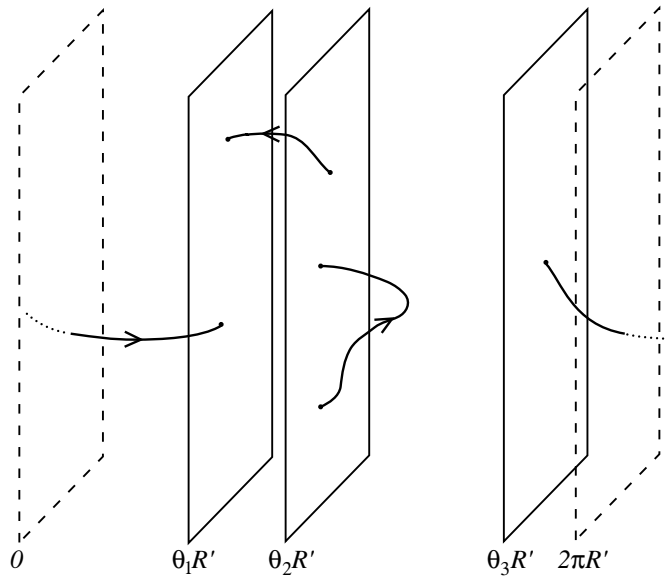}
\end{center}
\caption[]{$N=3$ hyperplanes at different positions, with various strings
attached.}
\end{figure}

\subsection{D-Brane Dynamics}

Let us first note that this whole picture goes through if several
coordinates $X^m=\{X^{25},X^{24},\ldots,X^{p+1}\}$ are periodic, and we
rewrite the periodic dimensions in terms of the dual coordinate.  The open
string endpoints are then confined to $N$ $(p+1)$-dimensional
hyperplanes.  The Neumann conditions on the world sheet, $\partial_n
X^m(\sigma^1,\sigma^2)=0$, have become Dirichlet conditions $\partial_t
X^{\prime m}(\sigma^1,\sigma^2)=0$ for the dual coordinates.

It is natural to expect that the hyperplane is dynamical rather than
rigid.\cite{dlp}  For one thing, this theory still has gravity, and
it is difficult to see how a perfectly rigid object could exist.
Rather, we would expect
that the hyperplanes can fluctuate in shape and position as dynamical
objects. We can see this by looking at the massless spectrum of the theory,
interpreted in the dual coordinates.

Taking for illustration the case where a single coordinate is dualized,
consider the mass spectrum.  The $D-1$ dimensional mass is
\bea
M^2 &=& (p^{25})^2+{1\over\ap}({N}-1) \nonumber\\
&=& \left( {[2\pi n+(\theta_i-\theta_j)]R^\prime\over2\pi\ap}
\right)^2 +{1\over\ap}(N-1).
\eea
Note that $[2\pi
n+(\theta_i-\theta_j)]R^\prime$ is the minimum length of a string winding
between hyperplanes $i$ and $j$.  Massless states arise
generically only for non-winding open strings whose end points are on the
same hyperplane, as the string tension contributes an energy to a stretched
string.  We have therefore the massless states:
\bea
\alpha^{\mu}_{-1}|{ k};ii\rangle, && V = \partial_t X^\mu, \nonumber\\
\alpha^{m}_{-1}|{ k};ii\rangle, && V = \partial_t X^{25} = \partial_n
X'^{25}.
\eea
The first of these is a gauge field
living on the hyperplane, with $p+1$ components tangent to the hyperplane. 
The second was the gauge field in the compact direction in the original
theory.  In the dual theory it becomes the transverse position of the
hyperplane.  We have already seen this in eq.~(\ref{dualpos}) for a Wilson
line, a constant gauge potential.  More general gauge backgrounds would
correspond to curved surfaces, and the quanta of the gauge fields to
fluctuations.  This is the same phenomenon as with spacetime itself.  We start
with strings in a flat background and discover that a massless closed
string state corresponds to fluctuations of the geometry.  Here we found
first a flat hyperplane, and then discovered that a certain open string
state corresponds to fluctuations of its shape.

Thus the hyperplane is indeed a dynamical object, a
Dirichlet membrane, or D-brane for short, or more specifically a D
$p$-brane.  In this terminology, the original Type I theory contains $N$
D 25-branes. A 25-brane fills space, so the string endpoint can be anywhere:
it just corresponds to an ordinary Chan-Paton factor.

It is interesting to look at the $U(N)$ symmetry breaking in the dual
picture.  When no D-branes coincide, there is just one massless vector
each, or $U(1)^N$ in all, the generic unbroken group.  If $m$
D-branes coincide, there are new massless states because
strings which are stretched between these branes
can achieve vanishing length.  Thus, there are $m^2$ vectors, forming
the adjoint of a $U(m)$ gauge group.\cite{joecomb}$^{\!,\,}$\cite{witbound}
This coincident position corresponds to 
$\theta_1=\theta_2=\cdots=\theta_m$ for some subset of the original
$\{\theta\}$, so in the original theory the Wilson line left a $U(m)$
subgroup unbroken. At the same time, there
appears a set of $m^2$ massless scalars: the $m$ positions are promoted
to a matrix.  This is curious and hard to visualize, but has proven to play
an important role in the dynamics of
D-branes.\cite{witbound}  Note that if all $N$
branes are coincident, we recover the $U(N)$ gauge symmetry.

This picture seems a bit exotic, and will become more so in the unoriented
theory.  But all we have done is to rewrite the original open string theory
in terms of variables which are more natural in the limit $R \ll
\sqrt{\ap}$.  Various puzzling features of the small-radius limit become
clear in the $T$-dual picture.

Observe that, since $T$-duality interchanges N and D boundary conditions, a
further $T$-duality in a direction tangent to a D $p$-brane reduces it to a
$(p-1)$-brane, while a $T$-duality in a direction orthogonal turns it into a
$(p+1)$-brane.  The case of a nontrivial angle will come up in the next
section.

\subsection{The D-Brane Action}

The world-brane
theory consists of a $U(1)$ vector field plus
$25-p$ world-brane scalars describing the fluctuations.  It is important to
consider the low energy effective action for these fields.
They are in
interaction with the massless closed string fields, whose action was given
in the first line of eq.~(\ref{stact}).
Introduce coordinates $\xi^a$, $a = 0, \ldots p$ on the brane.  The
fields on the brane are the embedding $X^\mu(\xi)$ and the gauge field
$A_a(\xi)$.  The action is\,\cite{leigh}
\be
S_p = -T_p \int d^{p+1}\xi\, e^{-\phi} \det\!^{1/2}\left(G_{ab} + B_{ab}
+ 2\pi\ap F_{ab}\right)\ , \label{dact}
\ee
where $G_{ab}$ and $B_{ab}$ are the pull-back of the spacetime fields to
the brane.  All features of this equation can be understood from general
reasoning.  The $\det\!^{1/2} G_{ab}$ term gives the world-volume. The
dilaton dependence $e^{-\phi} = g^{-1}$ arises because this is an open
string tree level action.

The dependence on $F$ can be understood as
follows.\cite{bachas}$^{\!,\,}$\cite{tdbi}  Consider a D-brane which is
extended in the $X^1$ and $X^2$ directions, with the other dimensions
unspecified, and let there be a constant gauge field
$F_{12}$. Go to the gauge $A_2 =X^1 F_{12}$.  Now $T$-dual along the
2-direction.  The open strings then satisfy Dirichlet conditions in this
direction, but the relation~(\ref{dualpos}) between the potential
and coordinate implies that the D-brane is tilted,
\be
X'^2 = 2\pi\ap X^1 F_{12}.
\ee
This gives a geometric factor in the action,
\be
\int dX^1\, \sqrt{1 + (\partial_1 X'^2)^2} = \int dX^1\, \sqrt{1 + (2\pi
\ap F_{12})^2}\ . \label{bitilt}
\ee
By boosting the D-brane to be aligned with the coordinate axes and then
rotating to bring $F_{ab}$ to block-diagonal form, one can reduce to a
product of factors~(\ref{bitilt}) giving $F_{ab}$ in the
determinant~(\ref{dact}). This determinant is the Born-Infeld
action.\cite{frad} The combination $B_{ab} + 2\pi\ap F_{ab}$ can be
understood as follows. These fields appear in the {string} world-sheet
action as
\be
\frac{1}{2\pi\ap} \int_{\cal M} B + \int_{\partial\cal M} A,
\ee
written using differential forms. This is invariant under the vector gauge
transformation $\delta A = d\lambda$, but the two-form gauge
transformation $\delta B = d\zeta$ gives a surface term that is canceled
by assigning a transformation $\delta A = -\zeta/2\pi\ap$ to the gauge
field.  Then the combination $B + 2\pi \ap F$ is invariant under both
symmetries, and it is this combination that must appear in the action.
Thus the form of the action is fully determined.

In the next section we will calculate the value of the tension $T_p$,
but it is interesting to note that one gets a recursion relation for it
from $T$-duality.\cite{tdtp}$^{\!,\,}$\cite{shanta}  The mass of a D $p$-brane
wrapped around a
$p$-torus is 
\be
T_p e^{-\phi} \prod_{i=1}^p (2\pi R_i)\ .\label{massp}
\ee  
Taking the $T$-dual on $X^p$ and recalling the
transformation~(\ref{tdil}) of the dilaton, we can rewrite the
mass~(\ref{massp}) in the dual variables:
\be
T_p (2\pi\sqrt{{\ap}}) e^{-\phi'} \prod_{i=1}^{p-1} (2\pi R_i)
\ =\ T_{p-1} e^{-\phi'} \prod_{i=1}^{p-1} (2\pi R_i)\ , 
\ee
or
\be
T_p = T_{p-1}/2\pi\sqrt{{\ap}}\ . \label{trec}
\ee 

For $N$ D-branes the brane fields become matrices as we have seen.
Non-derivative terms in the collective coordinates can be deduced by
$T$-duality from a constant $A_a$ field.  The leading action for the
latter, from the non-Abelian field strength, is proportional to
Tr$([X^a, X^b]\,[X_a,X_b])$.  The derivatives are relevant only when this
vanishes.  The $X_a$ then commute and can be diagonalized simultaneously,
giving $N$ independent D-branes.  The action is therefore
\bea
S_p &=& -T_p \int d^{p+1}\xi\, e^{-\phi} {\rm
Tr}\biggl\{ \det\!^{1/2}\left(G_{ab} + B_{ab} + 2\pi\ap F_{ab}\right)
\nonumber\\ 
&&\qquad\qquad\qquad + O([X^a, X^b]^2)
\biggr\}\ .
\label{ndact}
\eea
The precise form of the commutator term, including coupling to other fields
and higher powers of the commutator, can be deduced by $T$-duality from the
26-dimensional non-Abelian Born-Infeld action.

\subsection{D-Brane Tension}

It is instructive to compute the D-brane tension $T_{p}$, and for the
superstring the actual value will be significant.  As noted above, it is
proportional to
$g^{-1}$.  One could calculate it from the gravitational coupling to the
D-brane, given by the disk with a graviton vertex operator.  However, it
is much easier to obtain the absolute normalization as follows.
Consider two
parallel Dirichlet $p$-branes at positions $X^{\prime\mu}=0$ and
$X^{\prime\mu}=Y^\mu$. These two objects can feel each other`s presence by
exchanging closed strings as shown in figure~7.
\begin{figure}
\begin{center}
\leavevmode
\epsfbox{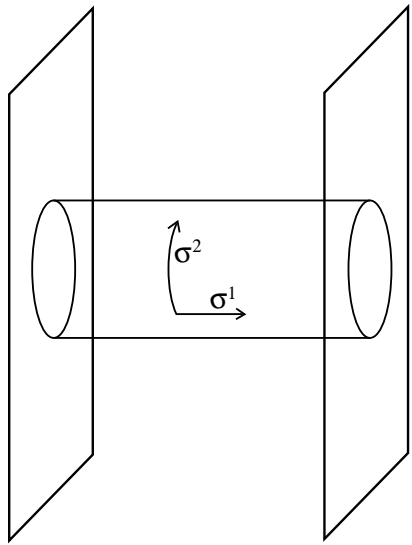}
\end{center}
\caption[]{Exchange of a closed string between two D-branes. 
Equivalently, a vacuum loop of an open string with one end on each
D-brane.}
\end{figure}
This string graph is an annulus, with no vertex operators.  
It is therefore easily calculated.  The poles from graviton and dilaton
exchange then give the coupling $T_p$ of closed string states to the
D-brane.

Parameterize the world-sheet as $(\sigma^1,\sigma^2)$ where
$\sigma^2$ (periodic) runs from $0$ to $2\pi t$, and $\sigma^1$ from $0$ to
$\pi$. This vacuum graph has the single modulus $t$, running from $0$ to
$\infty$.  Time-slicing horizontally, so that $\sigma^2$
is world-sheet time, gives a loop of open string.  Time-slicing
vertically instead, so that $\sigma^1$ is time, we see a single
closed string propagating in the tree channel.  The world-line of the open
string boundary can be regarded as a vertex connecting the vacuum to the
single closed string, i.e., a one-point closed string
vertex.\cite{chans}$^{\!,\,}$\cite{rrex}

Consider the limit $t \to 0$ of the loop amplitude. This is the 
ultra-violet limit for the open string channel, but unlike the torus, there
is no modular group acting to cut off the range of integration.  However,
because of duality, this limit is correctly interpreted as an {\it
infrared} limit.  Time-slicing vertically shows that the
$t \to 0$ limit is dominated by the lowest lying modes in the
closed string spectrum.  In keeping with string folklore, there are no
``ultraviolet limits'' of the moduli space  which could give rise to high
energy divergences.  All divergences in loop amplitudes come from pinching
handles and are controlled by the lightest states, or the long distance
physics.

One loop vacuum amplitudes are given by the Coleman-Weinberg
formula, which can be thought of as the sum of the zero point
energies of all the modes:\cite{cw}
\be
{\cal A} = V_{p+1}\int \frac{d^{p+1}k}{(2\pi)^{p+1}}\int_0^\infty {dt \over
2t} \sum_{I} e^{-2\pi \ap t (k^2 + M_I^2)} .
\ee
Here the sum $I$ is over the physical spectrum of the string, equivalent
to the transverse spectrum, and the momentum $k$ is in
the $p+1$ extended directions of the D-brane world-sheet.
The mass spectrum is given by
\be
M^2={1\over\ap}\left(\sum_{n=1}^\infty
n{\alpha}^i_{-n}\alpha_n^i-1\right)+{Y \cdot Y\over4\pi^2
\alpha^{\prime 2}}
\ee
where $Y^m$ is the separation of the D-branes.
The sums over $N_n^i = {\alpha}^i_{-n}\alpha_n^i$ are as usual
geometric, giving
\be
{\cal A}=2V_{p+1} \int_0^\infty {dt\over 2t} (8\pi^2 \ap t)^{-{(p+1)\over2}}
e^{-Y \cdot Y t/ 2\pi\ap} f_1(q)^{-24}\ .
\ee
Here $q=e^{-\pi t}$, the overall factor of 2 is from exchanging the
two ends of the string, and we define
\bea
f_1(q) &=& q^{1/12} \prod_{n=1}^{\infty} (1-q^{2n})
\nonumber\\
f_2(q) &=& \sqrt{2} q^{1/12} \prod_{n=1}^{\infty} (1+q^{2n})
\nonumber\\
f_3(q) &=& q^{-1/24} \prod_{n=1}^{\infty} (1+q^{2n-1})
\nonumber\\
f_4(q) &=& q^{-1/24} \prod_{n=1}^{\infty} (1-q^{2n-1})\ .
\eea  
We need the asymptotics as $t \to
0$.  The asymptotics as $t\to \infty$ are manifest, and the
$t\to 0$ asymptotics are then obtained from the modular transformations
\be
f_{1}(e^{-{\pi}/{s}}) = \sqrt{s}\,f_{1}(e^{-\pi s}),\quad
f_{3}(e^{-{\pi}/{s}}) = f_{3}(e^{-\pi s}),\quad
f_{2}(e^{-{\pi}/{s}}) = f_{4}(e^{-\pi s}) . 
\ee
In the present case
\be
{\cal A}=
2 V_{p+1} \int_0^\infty {dt\over 2t} (8\pi^2 \ap t)^{-{(p+1)\over2}}
e^{-Y \cdot Y t/ 2\pi\ap} t^{12} \left( e^{2\pi/t} + 24 + \ldots
\right).  \label{cpoles}
\ee
The leading divergence is from the tachyon and is an uninteresting bosonic
artifact.  The massless pole, from the second term, is
\bea
{\cal A} &\sim& V_{p+1}{24\over 2^{12}} (4\pi^2\ap)^{11-p}
\pi^{(p-23)/2} \Gamma((23 - p)/2) |Y|^{p-23} \nonumber\\
& =& V_{p+1} {24 \pi\over 2^{10}} (4\pi^2\ap)^{11-p}
G_{25-p}(Y^2)
\eea
where $G_d(Y^2)$ is the massless scalar Green's function in $d$ dimensions.

This can be compared with a field theory calculation, the exchange of
graviton plus dilaton between a pair of D-branes.  The propagator is from
the bulk action~(\ref{stact}) and the couplings are from the D-brane
action~(\ref{dact}).  This is a bit of effort because the graviton and
dilaton mix, but in the end one finds
\be
{\cal A} \sim {D-2 \over 4} V_{p+1} T_p^2 \kappa_0^2
 G_{25-p}(Y^2) \label{fieldpole}
\ee
and so
\be
T_p = {\sqrt{\pi}\over 16 \kappa_0} (4\pi^2\ap)^{(11-p)/2}\ .
\ee
This agrees with the recursion relation~(\ref{trec}).
The actual D-brane tension includes a factor of the dilaton from the
action~(\ref{dact}),
\be
\tau_p = {\sqrt{\pi}\over 16 \kappa} (4\pi^2\ap)^{(11-p)/2}\ 
\ee
where $\kappa = \kappa_0 e^\phi$.

As one application, consider $N$ 25-branes, which is the same as an ordinary
(fully Neumann)
$N$-valued Chan-Paton factor.  Expanding the 25-brane
Lagrangian~(\ref{dact}) to second order in the gauge field gives
\be
- {T_{25} \over 4} (2\pi\ap)^2 e^{-\phi}\Tr F_{\mu\nu}F^{\mu\nu},
\ee
with the trace in the fundamental representation of $U(N)$.  This gives the
precise numerical relation between the open and closed string
couplings.\cite{occoup} 

The asymptotics~(\ref{cpoles}) have an obvious interpretation in
terms of a sum over closed string states exchanged between the two D-branes.
One can write the cylinder path integral in Hilbert space formalism
treating
$\sigma_1$ rather than $\sigma_2$ as time.  It then has the form
\be
\langle B | e^{-(L_0 + \tilde L_0)\pi/t} | B \rangle
\ee
where the {\it
boundary state} $|B\rangle$ is the closed string state created by the
boundary loop.  We will not have time to develop
this formalism but it is useful in finding the couplings between
closed and open strings.\cite{chans}$^{\!,\,}$\cite{rrex}

\subsection{Unoriented Strings and Orientifolds.}

The $R \to 0$ limit of an unoriented theory also leads to new objects.
The effect of $T$-duality can easily be understood by viewing it as a
one-sided parity transformation.
For closed strings, the original coordinate is
$X^m(z,\zb)=X^m(z)+X^m(\zb)$ and the dual is $X^{\prime
m}(z,\zb)=X^m(z)-X^m(\zb)$.  The action of world sheet parity reversal is
to exchange $X^\mu(z)$ and $X^\mu(\zb)$. In terms of the dual coordinate
this is
\be
X^{\prime m}(z,\zb) \leftrightarrow -X^{\prime m}(\zb,z),
\ee
which is the product of a world-sheet {\it and} a spacetime parity operation.
In the unoriented theory, strings are invariant under the action of
$\Omega$, so in the dual theory we have gauged the product of world-sheet
parity with a spacetime symmetry, here parity.  This generalization of the
usual unoriented theory is known as an {\it orientifold,} a play on the term
orbifold for gauging a discrete spacetime symmetry.

Separate the string wavefunction into its internal
part and its dependence on the center of mass $x^m$, and take the internal
wavefunction to be an eigenstate of $\Omega$.  The projection then
determines the string wavefunction at $-x^m$ to be the same as at $x^m$, up
to a sign.  For example, the various components of the metric and
antisymmetric  tensor satisfy
\bea
G_{\mu\nu}(x^\mu,-x^m) = G_{\mu\nu}(x^\mu,x^m),&&
B_{\mu\nu}(x^\mu,-x^m) = -B_{\mu\nu}(x^\mu,x^m), \nonumber\\
G_{\mu n}(x^\mu,-x^m) = -G_{\mu n}(x^\mu,x^m),&&
B_{\mu n}(x^\mu,-x^m) = B_{\mu n}(x^\mu,x^m), \nonumber\\
G_{mn}(x^\mu,-x^m) = G_{mn}(x^\mu,x^m),&&
B_{mn}(x^\mu,-x^m) = -B_{mn}(x^\mu,x^m)\ . \\ \label{oriid}
\eea
In other words, the $T$-dual spacetime is the torus
$T^{25-p}$ modded by a $\Z_2$ reflection in the compact directions.
This is the same as the orbifold construction, the only difference
being the extra sign.
In the case of a single periodic
dimension, for example, the dual spacetime is the line segment $0 \leq x^{25}
\leq
\pi R'$,  with orientifold fixed planes at the ends.  It should be noted that
orientifold planes are not dynamical.  Unlike the case of D-branes, there are
no string modes tied to the orientifold plane to represent fluctuations in its
shape.\footnote{Our heuristic argument that a gravitational wave forces a
D-brane to oscillate does not apply to the orientifold fixed plane. 
Essentially, the identifications~(\ref{oriid}) become boundary conditions at
the fixed plane, such that the incident and reflected waves cancel.  For the
D-brane, the reflected wave is higher order in the string coupling.}  

Notice also that away from
the orientifold fixed planes, the local physics is that of the {\it
oriented} string theory.  Unlike the original unoriented theory, where the
projection removes half the states locally, here it relates the
amplitude to find a string at some point to the amplitude to find it at the
image point.

The orientifold construction was discovered via $T$-duality\,\cite{dlp} and
independently from other points of
view.\cite{opentwist}$^{\!,\,}$\cite{laterc}  One can of course consider more
general orientifolds which are not simply
$T$-duals of toroidal compactifications.  There have been quite a few papers
on orientifolds and their various duals.  This will not be a primary focus of
these lectures---we will be interested in orientifolds mainly insofar as they
arise in understanding the physics of D-branes.

In the case of open strings, the situation is similar.
Let us focus for convenience on a single compact dimension.  Again
there is one orientifold fixed plane at $0$ and another at $\pi R^\prime$.  
Introducing $SO(N)$ Chan-Paton factors, a Wilson line can be brought to
the form 
\be
{\rm diag}
\{\theta_1,-\theta_1,\theta_2,-\theta_2,\cdots,
\theta_{N/2},-\theta_{N/2}\}.
\ee
Thus in the dual picture there are 
${1\over 2}N$ D-branes on the line segment $0\leq X^{\prime25}<\pi
R^\prime$, and ${1\over 2}N$ at their image points under the orientifold
identification. 
Strings can stretch between D-branes and their images as shown.
\begin{figure}
\begin{center}
\leavevmode
\epsfbox{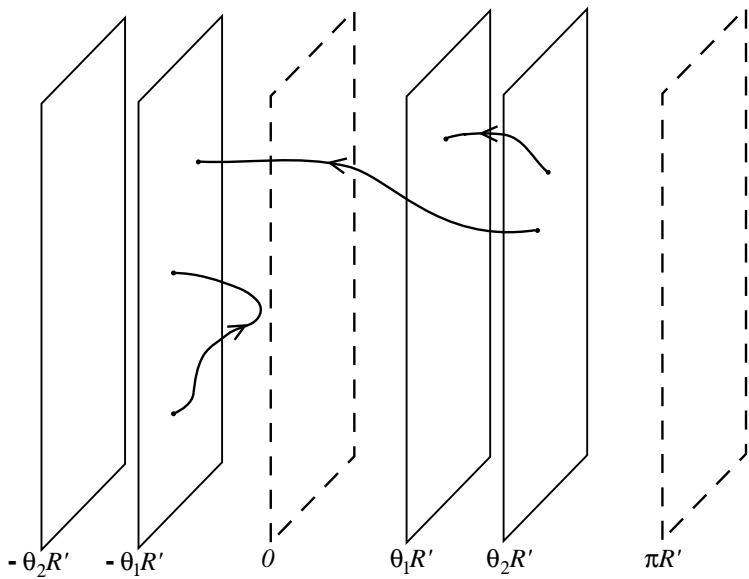}
\end{center}
\caption[]{Orientifold planes at $0$ and $\pi R'$, D-branes at
$\theta_1 R'$ and $\theta_2 R'$, and D-brane images at -$\theta_1 R'$ and
$-\theta_2 R'$.  The twist operator $\Omega$ acts on any string by a
combination of a spacetime reflection and reversing the orientation
arrow.}
\end{figure}
The generic gauge group is $U(1)^{N/2}$.  As in the oriented case,
if $m$ D-branes are coincident there is a $U(m)$ gauge group.
But now if the $m$ D-branes in addition lie at one of the fixed planes,
then strings stretching between one of these branes and one of the image
branes also become massless and we have the right spectrum of additional
states to fill out $SO(2m)$. The maximal $SO(N)$ is restored if all of
the branes are coincident at a
single orientifold plane.  Note that this maximally symmetric case is
asymmetric between the two fixed planes.
Similar considerations apply
to $USp(N)$.  

We should emphasize that there are $\frac{1}{2}N$
dynamical D-branes but an $N$-valued Chan-Paton index.  An interesting
case is when $m + \frac{1}{2}$ D-branes lie on a fixed plane, which makes
sense because the number $2m + 1$ of indices is integer.  A brane plus
image can move away from the fixed plane, but the number of branes remaining
is always half-integer.

The orientifold plane, like the D-brane, couples to the dilaton and
metric. The amplitude is the same as in the previous section, but with
$RP^2$ in place of the disk; that is, a crosscap replaces the boundary
loop.  The orientifold identifies $X^m$ with $-X^m$ at the opposite
point on the crosscap, so the crosscap is localized near one of the
orientifold fixed planes.  Again the easiest way to calculate this is
via vacuum graphs, the cylinder with one or both boundary loops replaced
by crosscaps.  These give the M\"obius strip and Klein bottle,
respectively.
\begin{figure}
\begin{center}
\leavevmode
\epsfbox{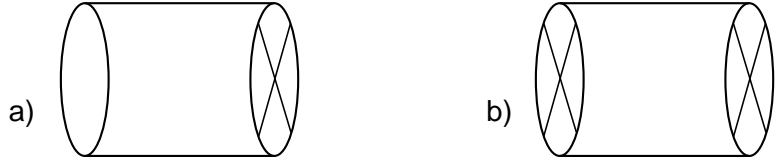}
\end{center}
\caption[]{a) M\"obius strip as cylinder with
crosscap at one end.  b) Klein bottle as cylinder with crosscaps at
each end.}
\end{figure}
To understand this, consider figure~10, which shows two copies of
the fundamental region for the M\"obius strip.
\begin{figure}
\begin{center}
\leavevmode
\epsfbox{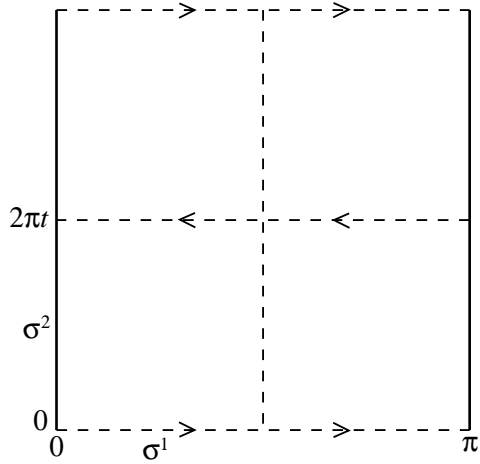}
\end{center}
\caption[]{Two copies of the fundamental region for the M\"obius strip.
}
\end{figure}
The lower half is identified with the reflection of the upper, and the
edges $\sigma^1 = 0, \pi$ are boundaries.  Taking the lower half as the
fundamental region gives the familiar representation of the M\"obius
strip as a strip of length $2\pi t$, with ends twisted and glued. 
Taking instead the left half of the figure, the line $\sigma^1 = 0$ is a
boundary loop while the line $\sigma^1 = \pi/2$ is identified with
itself under a shift
$\sigma^2 \to \sigma^2 + 2\pi t$ plus reflection of $\sigma^1$: it is a
crosscap.  The same construction applies to the Klein bottle, with the 
right and left edges now identified.

The M\"obius strip is given by the vacuum amplitude
\be
{\cal A}_{\rm M} = V_{p+1}\int \frac{d^{p+1}k}{(2\pi)^{p+1}} \int_0^\infty
{dt \over 2t} \sum_{I} {\Omega_i\over 2} e^{-2\pi \ap t (p^2 + M_I^2)} ,
\ee
where $\Omega_I$ is the $\Omega$ eigenvalue of state $i$.  The oscillator
contribution to $\Omega_I$ is $(-1)^N$ from eq.~(\ref{have}).\footnote
{In the directions orthogonal to the brane and orientifold there are two
additional signs in $\Omega_I$ which cancel: the world-sheet parity
contributes an extra minus sign in the directions with Dirichlet boundary
conditions (this is evident from the mode expansion~(\ref{modexps})), and the
spacetime reflection an additional sign.}  For
the $SO(N)$ open string the Chan-Paton factors have ${1\over 2}N(N+1)$ even
states and ${1\over 2}N(N-1)$ odd for a net of $+N$. For $USp(N)$ these
numbers are reversed for a net of $-N$.  Focus on a D-brane and its
image, which correspondingly contribute $\pm 2$.  The diagonal elements,
which contribute to the trace, are those where one end is on the D-brane
and one on its image.  The total separation is then $Y^m = 2X^m$.  Then,
\bea
{\cal A}_{\rm M} &=& \pm V_{p+1} \int_0^\infty {dt\over 2t} (8\pi^2 \ap
t)^{-{(p+1)\over2}} e^{-2X \cdot X t/ \pi\ap} \nonumber\\
&&\qquad\qquad\qquad\qquad\qquad
\left[ q^{-2} \prod_{k=1}^\infty (1+q^{4k-2})^{-24} (1-q^{4k})^{-24} \right]
\eea
The factor in braces $[\ ]$ is
\bea
f_3(q^2)^{-24} f_1(q^2)^{-24} &=& (2t)^{12} 
f_3(e^{-\pi/2t})^{-24} f_1(e^{-\pi/2t})^{-24} \nonumber\\
&=& (2t)^{12} \left(e^{\pi/2t} - 24 + \ldots \right)\ .
\eea
One thus finds a pole
\be
\mp 2^{p - 12} V_{p+1} {3 \pi\over 2^{6}}  (4\pi^2\ap)^{11-p}
G_{25-p}(X^2)\ .  \label{orten}
\ee
This is to be compared with the field theory result ${D-2\over2} V_{p+1}
T_p T'_p  \kappa_0^2 G_{25-p}(Y^2)$, where
$T'_p$ is the fixed-plane tension.  A factor of 2 as compared to the
earlier field theory calculation~(\ref{fieldpole}) comes because the
spacetime boundary forces all the flux in one direction.  Thus the fixed-plane
and D-brane tensions are related 
\be
\tau'_p = \mp 2^{p - 13} \tau_p.
\ee
A similar calculation with the Klein bottle gives a result proportional
to $\tau_p^{\prime 2}$.

Noting that there are $2^{25 - p}$ fixed planes, the total fixed-plane
source is $\mp 2^{12} \tau_p$.  The total source must vanish because the volume
is finite and there is no place for flux to go.  Thus there are $2^{12}$
D-branes (times two for the images) and the group is
$SO(2^{13}) = SO(2^{D/2})$.\cite{sobig} 
For this group the dilaton and graviton tadpoles cancel at order
$g^{-1}$.  This has no special significance in the bosonic string, as the
one loop $g^0$ tadpoles are nonzero and imaginary due to the tachyon
instability, but similar boundary combinatorics
will give a restriction on anomaly free Chan-Paton gauge groups in the
superstring.

\section{Lecture 3: Superstrings and $T$-Duality}

\subsection{Open Superstrings}

All of the exotic phenomena that we found in the bosonic string will
appear in the superstring as well, together with some important new
ingredients.  We first review open and unoriented superstrings.

The superstring world-sheet action is 
\be
S = {{1}\over{4 \pi}} \int_{\cal M} d^2\sigma  
  \{ \alpha^{\prime-1} \partial X^{\mu} \bar{\partial} X_{\mu}
     + \psi^{\mu} \bar{\partial} \psi_{\mu} + 
       \tilde{\psi}^{\mu}\partial \tilde{\psi}_{\mu} \}
\ee
where the open string world-sheet is the strip 
$0 < \sigma^1 < \pi$, $- \infty <\sigma^2< \infty$.
The condition that the surface term in the equation of 
motion vanishes allows two possible Lorentz invariant boundary conditions
on world-sheet fermions:
\bea
{\rm R\colon} && \psi^{\mu}(0, \sigma^2) =\tilde{\psi^{\mu}}
(0,\sigma^2) 
 \qquad\psi^{\mu}(\pi ,\sigma^2 )= \tilde{\psi^{\mu}}(\pi,\sigma^2)
\nonumber\\ 
{\rm NS\colon} && \psi^{\mu}(0, \sigma^2) =-\tilde{\psi^{\mu}}
(0,\sigma^2) 
\qquad\psi^{\mu}(\pi ,\sigma^2 )= \tilde{\psi^{\mu}}(\pi,\sigma^2)
\eea
We can always take the boundary condition at one end, say $\sigma^1 =
\pi$, to have a $+$ sign by redefinition of $\tilde\psi$.  The boundary
conditions and equations of motion are conveniently summarized by the {\it
doubling trick}, taking just left-moving (analytic) fields $\psi^\mu$ on
the  range $0$ to $2\pi$ and defining $\tilde\psi^\mu(\sigma^1,\sigma^2)$
to be $\psi^\mu(2\pi - \sigma^1,\sigma^2)$.
These left-moving fields are periodic in the Ramond (R) sector and
antiperiodic in the Neveu-Schwarz (NS).

In the NS sector the fermionic
oscillators are half-integer moded, giving a ground state energy of
$
(-{{8}\over{24}})+(-{{8}\over{48}})
 = - {1\over 2}
$
from the eight transverse coordinates and eight transverse
fermions.  The ground state is a
Lorentz singlet and has odd 
fermion number, $(-1)^F = -1$.  This assignment is necessary in order for
$(-1)^F$ to be multiplicatively conserved.\footnote{In the `$-1$
picture'\,\cite{fms}
the matter part of the ground state vertex operator is the
identity but the ghost part has odd fermion number.  In the `0 picture' this
is reversed.\label{picfoot}} The GSO projection, onto states with even fermion
number, removes the open string tachyon from the superstring spectrum. 
Massless particle states in ten dimensions are classified by their $SO(8)$
representation under Lorentz rotations that leave the momentum invariant. 
The lowest lying states in the NS sector are the eight transverse
polarizations of the massless open string photon, $A^{\mu}$, 
\be
\psi^{\mu}_{-1/2} |{ k}\rangle, \qquad M^2={{1}\over{\alpha'}} (N - \ha)
\ee
forming the vector of $SO(8)$. 

The fermionic oscillators in the Ramond sector are integer-moded.
In the R sector the ground state energy always vanishes
because the world-sheet bosons and their superconformal partners have the
same moding.
The Ramond vacuum is degenerate, since the $\psi^{\mu}_0$ take ground 
states into ground states, so the latter form a representation of the 
ten-dimensional Dirac matrix algebra  
\be
\{ \psi^{\mu}_0 , \psi^{\nu}_0 \} = \eta^{\mu \nu}\ .
\ee
The following basis for this representation is often convenient.  Form
the combinations
\bea
d^{\pm}_i &=& {{1}\over{\sqrt 2}}\left ( \psi^{2i}_0\pm i \psi^{2i+1}_0\right
) \qquad i=1,\cdots,4 \nonumber\\ 
 d^{\pm}_0 &=& {{1}\over{\sqrt 2}}\left ( \psi^{1}_0 \mp \psi^{0}_0\right ) 
\eea
In this basis, the Clifford algebra takes the form
\be
\{ d^{+}_i, d^{-}_j \}=\delta_{ij}\ .
\ee
The $d^{\pm}_i$, $i = 0, \cdots, 4$ act as raising and lowering 
operators, generating the 32 Ramond ground states.  Denote these
states
\be
|s_0,s_1,s_2,s_3,s_4 \rangle = |{\bf s}\rangle
\ee
where each of the $s_i$ is $\pm\ha$, and where
\be
d^{-}_{i} | -\ha , -\ha , -\ha , -\ha , -\ha \rangle = 0  
\ee
while $d^{+}_i$ raises $s_i$ from $-\ha$ to $\ha$.
The significance of this notation is as follows.  The fermionic part of the
ten-dimensional Lorentz generators is 
\be
S^{\mu \nu} = - {{i}\over{2}} \sum_{r \in {\bf Z }+\kappa} 
  \psi^{[ \mu}_{-r}\psi^{\nu ]}_r  
\ee
where $\kappa=0$ ($\ha$) in the R (NS) sector.  The states above are
eigenstates of $S_0 = iS^{01}$, $S_i = S^{2i,2i+1}$, with $s_i$ the
corresponding eigenvalues.  Since the Lorentz generators always flip an
even number of $s_i$, the Dirac representation $\bf 32$ decomposes into a
$\bf 16$ with an even number of $-\ha$'s and $\bf 16'$ with an odd number.

Physical states are annihilated by the zero mode of the supersymmetry 
generator, which on the ground states
reduces to $G_0=p_{\mu}\psi^{\mu}_0$. In the frame $p^0 = p^1$ this
becomes $s_0 = \ha$,
giving a sixteen-fold degeneracy for the {\it physical} 
Ramond vacuum.  This is a representation of $SO(8)$ which again
decomposes into ${\bf 8_s}$ with an even number of $-\ha$'s and ${\bf
8_c}$ with an odd number.

The GSO projection keeps one irreducible representation; the two choices,
$\bf 16$ or $\bf 16'$, are physically equivalent, differing only by a
spacetime parity redefinition.
It is useful
to think of the GSO projection in terms of locality of the OPE of a general
vertex operator with the gravitino vertex operator.  Suppose we take a
projection which includes the operator $e^{-\phi/2 +
i(H_0+H_1+H_2+H_3+H_4)/2}$, where the
$H_i$ are the bosonization of $\psi^\mu$.\cite{fms}  In
the NS sector this has a branch cut with the ground state vertex operator
$e^{-\phi}$, accounting for the sign discussed in footnote~{\it\ref{picfoot}}
for the
$-1$ picture vertex operator.  In the R sector the ghost plus longitudinal
part of the OPE is local, so we have
\be
\sum_{i=1}^4 s_i = 0 \pmod 2,
\ee
picking out the ${\bf 8_s}$.

The ground state spectrum is then ${\bf 8}_v \oplus {\bf 8}_s$, a vector
multiplet of $D=10$, $N=1$ spacetime supersymmetry.  Including Chan-Paton
factors gives again a $U(N)$ gauge theory in the oriented theory and
$SO(N)$ or $USp(N)$ in the unoriented.

\subsection{Closed Superstrings}
The closed string spectrum is the product of two copies of the open
string spectrum, with right- and left-moving levels matched.
In the open string the two choices for the GSO projection were
equivalent, but in the closed string there are two inequivalent choices,
taking the same (IIB) or opposite (IIA) projections on the two sides.
These lead to the massless sectors
\bea
{\rm Type ~ IIA\colon} && ({\bf 8_v}\oplus{\bf 8_s}) \otimes
   ({\bf 8_v}\oplus{\bf 8_{c}}) \nonumber\\
{\rm Type ~ IIB\colon} && ({\bf 8_v}\oplus{\bf 8_s}) \otimes
   ({\bf 8_v}\oplus{\bf 8_{s}}) 
\eea
of $SO(8)$.  

The various products are as follows.  In the NS-NS sector,
this is
\be
{\bf 8_v} \otimes {\bf 8_v} = \phi \oplus B_{\mu\nu} \oplus G_{\mu\nu}
={\bf 1} \oplus {\bf 28}  \oplus {\bf 35} .
\ee
In the R-R sector, the IIA and IIB spectra are respectively
\bea
{\bf 8_s} \otimes {\bf 8_c} &=& [1] \oplus [3] = {\bf 8_v} \oplus
{\bf 56_t} \nonumber\\
{\bf 8_s} \otimes {\bf 8_s} &=& [0] \oplus [2] \oplus [4]_+
= {\bf 1} \oplus {\bf 28}  \oplus {\bf 35}_+ .
\eea
Here $[n]$ denotes the $n$-times antisymmetrized representation of
$SO(8)$, with $[4]_+$ being self-dual.  Note that the representations
$[n]$ and $[8-n]$ are the same, being related by contraction with the
8-dimensional $\epsilon$-tensor.
The NS-NS and R-R spectra
together form the bosonic components of $D=10$ IIA (nonchiral) and IIB
(chiral) supergravity respectively.  In the NS-R and R-NS sectors are
the products
\bea
{\bf 8_v} \otimes {\bf 8_c} &=& 
{\bf 8_s}\oplus{\bf 56_c} \nonumber\\
{\bf 8_v} \otimes {\bf 8_s} &=&
{\bf 8_c}\oplus{\bf 56_s}.
\eea
The $\bf 56_{s,c}$ are gravitinos, their vertex operators having one
vector and one spinor index.  They must couple to conserved spacetime
supercurrents.  In the IIA theory the two gravitinos (and
supercharges) have opposite chirality, and in the IIB the same.

Let us develop further the vertex operators for the R-R states.  
This will involve a product of spin fields,\cite{fms}
$e^{-{{\varphi}\over{2}} - {{\tilde{\varphi}}\over{2}} } S_{\alpha}
\tilde{S}_{\beta}$.  These again decompose into antisymmetric tensors, now
of $SO(9,1)$: 
\be
V = e^{ -{{\varphi}\over{2}} - {{\tilde{\varphi}}\over{2}} } S_{\alpha} 
\tilde{S}_{\beta}
( \Gamma^{[\mu_1} \cdots \Gamma^{\mu_n]}C)_{\alpha\beta} G_{[\mu_1 \cdots
\mu_n]}(X) \label{rrver}
\ee
with $C$ the charge conjugation matrix.  In the IIA theory the
product is ${\bf 16} \otimes {\bf 16'}$ giving even $n$ (with $n \cong
10-n$) and in the IIB theory it is ${\bf 16} \otimes {\bf 16}$ giving odd
$n$.
As is usual, the classical equations of motion follow from the 
physical state conditions, which at the massless level reduce to
$G_0 \cdot V = \tilde{G}_0 \cdot V = 0.$
The relevant part of $G_0$ is just $p_\mu \psi^\mu_0$ and similarly for
$\tilde G_0$.  The $p_\mu$ acts by differentiation on $G$, while
$\psi_0^\mu$ acts on the spin fields as it does on the corresponding
ground states: as multiplication by $\Gamma^\mu$.  Noting the identity
\be
\Gamma^\nu \Gamma^{[\mu_1} \cdots \Gamma^{\mu_n]} =
\Gamma^{[\nu} \cdots \Gamma^{\mu_n]} +
\left( \delta^{\nu\mu_1} \Gamma^{[\mu_2} \cdots \Gamma^{\mu_n]}
+ {\rm perms} \right)  \label{gamma}
\ee
and similarly for right multiplication, the physical state conditions
become
\be
dG=0 \qquad\qquad d{}^* G = 0.
\ee
These are the Bianchi identity and field equation for an antisymmetric
tensor field strength.  This is in accord with the representations found:
in the IIA theory we have odd-rank tensors of $SO(8)$ but even-rank
tensors of $SO(9,1)$ (and reversed in the IIB), the extra index being
contracted with the momentum to form the field strength.
It also follows that R-R amplitudes involving elementary strings vanish
at zero momentum, so strings do not carry R-R charges.

As an aside, when the dilaton background is nontrivial, the Ramond
generators have a term $\phi_{,\mu} \partial\psi^\mu$, and the Bianchi
identity and field strength pick up terms proportional to
$d\phi \wedge G$ and $d\phi \wedge {}^*G$.  The Bianchi identity is
nonstandard, so $G$ is not of the form $dC$.  Defining $G' = e^{-\phi} G$
removes the extra term from both the Bianchi identity and field strength.
The field $G'$ is thus decoupled from the dilaton.
In terms of the action, the fields $G$ in the vertex operators appear
with the usual closed string $e^{-2\phi}$ but with non-standard dilaton
gradient terms.  The fields we are calling $G'$ (which in fact are the
usual fields used in the literature) have a dilaton-independent action.

The IIB theory is invariant under world-sheet parity, so we can again
form an unoriented theory by gauging.  
Projecting onto $\Omega =+1$ interchanges left-moving and right-moving 
oscillators and so one linear combination of the R-NS and NS-R gravitinos
survives, leaving $D=10$, $N=1$ supergravity.  In the NS-NS sector, the
dilaton and graviton are symmetric under $\Omega$ and survive, while the
antisymmetric tensor is odd and is projected out. In the R-R sector, it
is clear by counting that the $\bf 1$ and ${\bf 35}_+$ are in the
symmetric product of ${\bf 8_s} \otimes {\bf 8_s}$ while the $\bf 28$ is
in the antisymmetric.  The R-R vertex operator is the product of right-
and left-moving fermions, so there is an extra minus in the exchange and
it is the $\bf 28$ that survives.  The bosonic massless sector is
thus ${\bf 1} \oplus {\bf 28}  \oplus {\bf 35}$, the $D=10$ $N=1$
supergravity multiplet.  This is the same multiplet as in the heterotic
string, but now the antisymmetric tensor is from the R-R sector.

The open superstring has only $N=1$ supersymmetry.  In order that the
closed strings couple consistently they must also have $N=1$ supergravity
and so the theory must be unoriented.  In fact, spacetime anomaly
cancellation implies that the only consistent $N=1$ superstring is the
$SO(32)$ open plus closed string theory.  Now, as a general principle any
such inconsistency in the low energy theory should be related to some stringy
inconsistency.  This is the case, but it will be more convenient to
discuss this later after some discussion of $T$-duality.

\subsection{$T$-Duality of Type II Superstrings}

Even in the closed oriented Type II theories $T$-duality has an
interesting effect.\cite{dhs}$^{\!,\,}$\cite{dlp}.  Consider
compactifying a single coordinate $X^9$.  In the $R\to \infty$ limit the
momenta are $p^9_R = p^9_L$, while in the $R \to 0$ limit $p^9_R = -p^9_L$. 
Both theories are
$SO(9,1)$ invariant but under {\it different} $SO(9,1)$'s.
Duality reverses the sign of the right-moving $X^9(\zb)$; therefore by
superconformal invariance it does so on $\tilde\psi^9(\zb)$.  Separate the
Lorentz generators into their left-and right-moving parts $M^{\mu\nu} +
\tilde M^{\mu\nu}$. Duality reverses all terms in $\tilde M^{\mu 9}$, so the
$\mu 9$ Lorentz generators of the $T$-dual theory are $M^{\mu 9} - \tilde
M^{\mu 9}$. In particular this reverses the sign of the helicity $\tilde
s_4$ and so switches the chirality on the right-moving side.  If one starts
in the IIA theory, with opposite chiralities, the $R\to 0$ theory has the
same chirality on both sides and is the $R\to\infty$ limit of the IIB theory,
and vice versa.  More simply put, duality is a one-sided spacetime parity
operation and so reverses the relative chiralities of the right- and
left-moving ground states.  The same is true if one dualizes on any odd
number of dimensions, while dualizing on an even number returns the original
Type II theory.

Since the IIA and IIB theories have different R-R fields, $T_9$ duality
must transform one set into the other.  The action of duality on the spin
fields is of the form
\be
S_{\alpha} (z) \to S_{\alpha} (z),\qquad
\tilde{S}_{\alpha} (\bar{z}) \to P_9 \tilde{S}_{\alpha} (\bar{z}) 
\ee
for some matrix $P_9$, the parity transformation (9-reflection) on the
spinors.  In order for this to be consistent with the action $\tilde\psi^9 \to
-\tilde\psi^9$,
$P_9$ must anticommute with
$\Gamma^9$ and commute with the remaining $\Gamma^\mu$.  Thus
$P_9 =
\Gamma^9\Gamma^{11}$ (the phase of $P_9$ is determined, up to sign, by
hermiticity of the spin field).  Now consider the effect on the R-R vertex
operators~(\ref{rrver}).  The $\Gamma^{11}$ just contributes a sign, because
the spin fields have definite chirality.  Then by the $\Gamma$-matrix
identity~(\ref{gamma}), the effect is to add a 9-index to $G$ if none is
present, or to remove one if it is.  The effect on the potential $C$ ($G =
dC$) is the same.  Take as an example the Type IIA vector $C_\mu$.
The component $C_9$ maps to the IIB scalar $C$, while the $\mu\neq 9$
components map to $C_{\mu 9}$.  The remaining components of $C_{\mu\nu}$
come from $C_{\mu \nu 9}$, and so on.

\subsection{$T$-Duality of Type I Superstrings}

The action of $T$-duality in the open and unoriented Type I theory
produces D-branes and orientifold planes, just as in the bosonic string. 
Let us focus here on a single D-brane, taking a limit
in which the other D-branes and the orientifold planes are
distant and can be ignored.  Away from the D-brane, only closed strings
propagate.  The local physics is that of the Type II theory, with two
gravitinos.  This is true even if though we began with the unoriented Type
I theory which has only a single gravitino.  The point is that the closed
string begins with two gravitinos, one with the spacetime
supersymmetry on the right-moving side of the world-sheet and one on the
left.  The orientation projection of the Type I theory leaves one linear
combination of these.  But in the $T$-dual theory, the
orientation projection does not constrain the local state of the string,
but relates it to the state of the (distant) image gravitino.
Locally there are two independent gravitinos, with equal chiralities if an
even number of dimensions have been dualized and opposite if an odd number.

However, the open string boundary conditions are invariant under only one
supersymmetry.  In the original Type I theory, the left-moving world-sheet
current for spacetime supersymmetry $j_\alpha(z)$ flows into the boundary
and the right-moving current $\tilde j_\alpha(\bar z)$ flows out, so
only the total charge $Q_\alpha + \tilde Q_\alpha$ of the
left- and right-movers is conserved.  Under $T$-duality this becomes
\be
Q_\alpha + \left({\textstyle \prod_m} P_m\right) \tilde Q_\alpha\ , 
\ee
where the product of reflections $P_m$ runs over
all the dualized dimensions, that is, over all directions orthogonal to the
D-brane.  Closed strings couple to open, so the general
amplitude has only one linearly realized supersymmetry.  That is, the vacuum
without D-branes is invariant under
$N=2$ supersymmetry, but the state containing the D-brane is invariant under
only $N=1$: {\it it is a BPS state.}\cite{joeone}$^{\!,\,}$\cite{gdinst}

BPS states must carry conserved charges.  In the present case there is
only one set of charges with the correct Lorentz properties, namely the
antisymmetric R-R charges.  The world volume of a $p$-brane naturally
couples to a ($p + 1$)-form potential $C_{(p+1)}$, which has a
($p + 2$)-form field strength $G_{(p+2)}$.  This identification can also
be made from the $g^{-1}$ behavior of the D-brane tension: this is the
behavior of an R-R soliton.\cite{blackp}$^{\!-\,}$\cite{coni}, as will be
developed further in section~5.8.

The IIA theory has $p = 0$, 2, 4, 6, and 8-branes.  The vertex
operators~(\ref{rrver}) describe field strengths of all even ranks from 0
to 10.  By a
$\Gamma$-matrix identity the $n$-form and $(10-n)$-form field strengths
are Hodge dual to one another, so a $p$-brane and $(6-p)$-brane are
sources for the same field, but one `magnetic' and one `electric.'  The field
equation for the 10-form field strength allows no propagating states, but
the field can still have a physically significant energy
density~\cite{joeone}$^{\!,\,}$\cite{romans}$^{\!-\,}$\cite{joeandy}. 
Curiously, the 0-form field strength should couple to a $(-2)$-brane, but it
is not clear how to interpret this---perhaps there is something interesting
to learn here. 

The IIB theory has $p = -1$, 1, 3, 5, 7, and 9-branes.  The vertex 
operators~(\ref{rrver}) describe field strengths of all odd ranks from 1 to
9, appropriate to couple to all but the 9-brane.  The 9-brane does couple to
a nontrivial {\it potential,} as we will see below.

A $(-1)$-brane is a
Dirichlet instanton, defined by Dirichlet conditions in the time direction
as well as all spatial directions.\cite{parton}  Of course, it is not clear
that
$T$-duality in the time direction has any meaning, but one can argue for
the presence of $(-1)$-branes as follows.  Given $0$-branes in the IIA
theory, there should be virtual $0$-brane world-lines that wind in a
purely spatial direction.  Such world-lines are required by quantum
mechanics, but note that they are essentially instantons, being localized
in time.  A $T$-duality in the winding direction then gives a $(-1)$-brane.
One of the first clues to the relevance of D-branes\,\cite{joecomb} was the
observation that D-instantons, having action $g^{-1}$, would contribute
effects of order $e^{-1/g}$ as expected from the behavior of large orders of
string perturbation theory.\cite{shenk1}

\subsection{The D-Brane Action and Charge}

We have concluded that the D-brane must couple to a ($p + 1$)-form
potential.  The spacetime plus D-brane action
then includes
\be
S={1\over2}\int G_{(p+2)}{}^*G_{(p+2)}+i\mu_p\int_{p-\rm brane} C_{(p+1)},
\label{formaction}
\ee
where the ($p + 1$)-form charge of the D $p$-brane is
denoted $\mu_p$.\footnote{This action is not correct for $p=3$, for which
the field strength is self-dual. There is no covariant action in this case.}
As discussed earlier, the dilaton does not appear in the action.  However,
there are additional terms involving the D-brane gauge field, similar to
the Born-Infeld terms.  Again these can be determined from $T$-duality.
Consider, as an example, a 1-brane in the 1-2 plane.  The action is
\be
\int dx^1\, \left( C_1 + \partial_1 X^2 C_2\right)\ .
\ee
Under a $T$-duality in the 2-direction this becomes
\be
\int dx^1\, \left( C_{12} + 2\pi\apm F_{12} C
\right)\ . \label{rrf}
\ee
We have used the $T$-transformation of the $C$ fields as discussed in
section~3.3.\footnote{We have not kept track of the overall normalization but
one could, with the result~$\mu_p = \mu_{p-1}/2\pi\sqrt{{\ap}}$ analogous to
the earlier result~(\ref{trec}) for $T_p$.  Instead we will obtain $\mu_p$
directly later.}  This argument is easily generalized, with the
Chern-Simons-like result
\be
i\mu_p\int_{p-\rm brane}  {\rm Tr}\left(e^{2\pi\apm F + B}\right)
\textstyle{\sum}_p C_{(p+1)} \ . \label{csact}
\ee
The expansion of the integrand~(\ref{csact}) involves forms of various rank;
the integral picks out precisely the terms
that are proportional to the volume form of the
$p$-brane.\cite{rract}$^{\!-\,}$\cite{douginst}$
^{\!,\,}$\cite{tdbi}$^{\!,\,}$\cite{tdtp}  There are also couplings of the R-R
potentials to curvature.\cite{bsv}$^{\!,\,}$\cite{ghs}  In addition, the
coupling of the D-brane to NS-NS and open string states has the same
form~(\ref{dact}) as the bosonic D-brane theory.

The D-brane, unlike the fundamental string, carries R-R charge.  It
is interesting to see how this is consistent with our earlier discussion
of string vertex operators.\cite{bpspict}
The R-R vertex operator~(\ref{rrver}) is in the
$(-\ha ,-\ha)$ picture, which can be used in almost all processes.  In
the disk, however, the total right+left ghost number must be $-2$.  With
two or more R-R vertex operators, all can be in the $(-\ha ,-\ha)$
picture (with picture changing operators included as well), but a single
vertex operator must be in either the $(-{3\over 2}, -\ha)$ or the
$(-\ha,-{3\over 2})$ picture.  The  $(-\ha ,-\ha)$ vertex operator is
essentially $e^{-\varphi} G_0$ times the $(-{3\over 2}, -\ha)$ operator,
so besides the shift in the ghost number the latter has one less power of
momentum and one less $\Gamma$-matrix. The missing factor of momentum
turns $G$ into $C$, and the missing $\Gamma$-matrix gives the correct
Lorentz representations for the potential rather than the field strength.

To obtain the D-brane tension and R-R charge, one can consider the same
vacuum cylinder as in the bosonic string.\cite{joeone}
Carrying out the traces over the open superstring spectrum
gives
\bea
A &=& 2V_{p+1} \int {dt\over 2t}\, (8\pi^2 \ap t)^{-(p+1)/2}
e^{- t{Y^2\over2\pi \alpha'}} 
\nonumber\\
&& \qquad\qquad\qquad\qquad\qquad \frac{ -f_2(q)^{8}
+ f_3(q)^8 - f_4(q)^8 }{2 f_1^8(q)},
\eea
where again $q =e^{-\pi t}$.
The three terms in the fraction come from the open string R
sector with ${1\over2}$ in the trace, from the NS sector with
${1\over2}$ in the trace, and the NS sector with ${1\over2} (-1)^F$
in the trace; the R sector with ${1\over2} (-1)^F$ gives no net
contribution.  These three terms sum to zero by the `abstruse
identity,' because the open string spectrum is supersymmetric.
In terms of the closed string exchange, this
reflects the fact that D-branes are BPS states, the net forces from
NS-NS and R-R exchanges canceling.  The separate exchanges can be
identified as follows.  In the terms with $(-1)^F$, the world-sheet
fermions are periodic around the cylinder thus corresponding to R-R
exchange, while the terms without $(-1)^F$ have antiperiodic fermions and
come from NS-NS exchange.  Obtaining the $t\to 0$ behavior as before
gives
\bea
{\cal A}_{\rm NS}\ =\ - {\cal A}_{\rm R} 
 &\sim & {1\over2} V_{p+1}  \int{dt\over t} (2\pi
t)^{-(p+1)/2} (t/2\pi\alpha')^4 e^{- t{Y^2\over8\pi^2 \alpha'^2}} 
\nonumber\\
&=& V_{p+1} 2\pi (4\pi^2\alpha')^{3-p} G_{9-p}(Y^2).
\eea 
Comparing with field theory calculations gives\,\cite{joeone}
\be
\mu_p^2 = 2 \kappa^2 \tau_p^2 = 2\pi (4\pi^2\alpha')^{3-p}. \label{dcharge}
\ee

D-branes that are not parallel feel a
net force because the cancellation is no longer exact.  In the extreme case,
where one of the D-branes is rotated by $\pi$, the coupling to the dilaton
and graviton is unchanged but the coupling to the R-R tensor is reversed in
sign, and the two terms in the cylinder amplitude add.  In fact, a
well-known divergence of Dirichlet boundary conditions sets in for
non-parallel branes: the
$t$-integration diverges at zero.  This is similar to the Hagedorn
divergence, and represents an instability of antiparallel D-branes when brought
too close.\cite{tomlen}

The orientifold planes also break half the supersymmetry and are R-R and
NS-NS sources.  In the original Type I theory the orientation projection
keeps only the linear combination $Q_\alpha + \tilde
Q_\alpha$.  In the $T$-dualized theory this becomes $Q_\alpha + (\prod_m
P_m)
\tilde Q_\alpha$ just as for the D-branes.  The force between an
orientifold plane and a D-brane can be obtained from the M\"obius strip as
in the bosonic case; again the total is zero and can be separated into
NS-NS and R-R exchanges.  The result is similar to the bosonic
result~(\ref{orten}),
\be
\mu'_p = \mp 2^{p - 5} \mu_p, \qquad \tau'_p = \mp 2^{p - 5} \tau_p \ .
\ee
Since there are $2^{9-p}$ orientifold planes, the total fixed-plane charge
is $\mp 16 \mu_p$, and the total fixed-plane tension is $\mp 16 \tau_p$.

A nonzero total tension represents a source for the graviton and dilaton,
so that at order $g$ these fields become time dependent as in the
Fischler-Susskind mechanism.\cite{fsuss}
A nonzero total R-R source is more serious: the field equations are
inconsistent, because R-R flux lines have no place to go in the compact
space.\footnote
{The Chern-Simons coupling~(\ref{csact}) implies that the open
string field strengths are also R-R sources, so there can also be
consistent solutions with nonzero values for these.}
So we need exactly 16
D-branes with the $SO$ projection, giving the $T$-dual of $SO(32)$.  

The spacetime anomalies for $G \neq SO(32)$ are thus accompanied by a
divergence.\cite{gsdiv}  The reason for this is as follows.  As in field
theory, one can relate the anomaly to the ultraviolet limit of an (open
string) loop graph.  But string theory has no true ultraviolet limits.  This
limit of the annulus ($t\to \infty$) is instead the infrared limit of the
closed string tree graph, and the anomaly comes from this infrared
divergence.  From the 
world-sheet point of view there is a conformal anomaly that cannot be
canceled because of the inconsistency of the field equations.  All this
applies even in the original $D=10$ Type I theory.\cite{rrex}  The Neumann
open strings correspond to 9-branes, since the endpoints can be anywhere. 
The Dirichlet and orientifold 9-branes couple to an R-R 10-form,
\be
i (32 \mp N) {\mu_{10}\over 2} \int A_{10} ,
\ee
and the field equation from varying~$A_{10}$ is just $G =
SO(32)$!\,\cite{rrex}

\section{Lecture 4: Advanced D-Mechanics}

\subsection{Discussion}

We have seen that $T$-duality of the Type I string leads to a theory with
precisely 16 Dirichlet $p$-branes on a $T_{9-p}/Z_2$ orientifold, for any
given value of $p$.
We now understand that the restriction to 16 comes
from conservation of R-R charge.  It follows that in a non-compact space,
where the flux lines could run to infinity, we could have a consistent
theory with any number and configuration of $p$-branes, with all $p$
being even in the IIA theory or odd in the IIB.  

Indeed, cluster
decomposition plus $T$-duality forces this upon us.  Let us start from the
Type I theory and see how far we can get by combination of $T$-dualities,
turning on background fields, taking limits, and cluster decomposition.
First, let us take the $T$-dual on all spatial directions and then the
$R\to 0$ limit.  We are left with 16 D 0-branes at arbitrary positions. 
But now cluster decomposition says\,\footnote
{This is my paraphrase.  See Weinberg's text\,\cite{weinberg} for a more
discussion.} that if we can have a state with 16 0-branes in this room and
none behind the moon, or vice versa, then we can also have a state with 16
in each place, or none.  That is, states with arbitrary numbers of 0-branes
are in the spectrum.  In particular, the state with no 0-branes is just
what we would call the Type~IIA theory, giving also the IIB theory by
$T$-duality.  So both Type II theories can be regarded as states within the
Type I theory.

Similarly by $T$-duality from this we get a state with any number of
$p$-branes for any fixed $p$.  The $T$-dual of a
flat torus gives flat D-branes, but because they are dynamical this is
continuously connected to configurations where the D-branes fold back and
forth, and in this way one can reach a configuration which in any local
region has an arbitrary set of $p$-branes.  Now rotate some of the
$p$-branes
$90^\circ$ (which is $T$-dual to turning on a gauge field) and 
$T$-dualize in a direction parallel to some and perpendicular to others. 
Some become $(p+1)$-branes and some $(p-1)$-branes.  In this way we can
obtain any collection of even D-branes in the IIA theory or odd D-branes
in the IIB.  So far, these D-branes are all wound around tori, or infinite
in the limit, but again by a cluster-type argument one should be able to
build a D-brane of arbitrary topology out of D-stuff that is locally the
same.  So what is usually called the Type~I theory should be regarded as
including the Types~I, IIA and IIB theories with any collection of D-branes.
These are all consistent string theories, provided that the D-branes
satisfy the appropriate equations of motion and the R-R field equations
are consistent. 

It is conceivable that there is another Type II string theory that has no
D-branes at all, but Occam's razor would suggest that there is likely to be
only one non-perturbative completion of a given string theory.

Now that we are considering configurations of $p$-branes with several values
of $p$ there is an important consistency check.  The field strengths to
which a $p$-brane and $(6 - p)$-brane couple are dual to one another,
$G_{(p+2)} = *G_{(8-p)}$.  This implies a Dirac quantization condition, as
generalized by Teitelboim and Nepomechie.\cite{dirac}  Integrating the field
strength $*G_{(p+2)}$ on an ($8 - p$)-sphere surrounding a $p$-brane,
the action~(\ref{formaction}) gives a total flux $\Phi=
\mu_p$.  We can write
$*G_{(p+2)} = G_{(8-p)} = d C_{(7-p)}$ everywhere except on a Dirac
`string'.  Then
\be
\Phi= \int_{S_{8-p}} *G_{(p+2)}=\int_{S_{7-p}} C_{(7-p)}\ .
\ee
where we perform the last integral on a small sphere surrounding the Dirac 
string.  A ($6 - p$)-brane circling the string picks up a phase
$e^{i \mu_{6 - p}\Phi}$.  The condition that the string be invisible is
\be
\mu_{6 - p} \Phi = \mu_{6 - p} \mu_p = 2\pi n.
\ee
The D-branes charges~(\ref{dcharge}) satisfy this with the minimum
quantum $n=1$.\footnote
{This argument does not apply directly to the case $p=3$, as the self-dual
5-form field strength has no covariant action, but the result follows for
$p = 3$ by $T$-duality.  Consider 3-branes extended in the 456 and 789
directions.  Both are local in the 123 direction.  The Dirac quantization
argument is made by considering the dependence of the wavefunction on the 123
coordinate; but this is the same if we $T$-dualize to a 45 2-brane and a
6789 4-brane, for which $n=1$ is already established.}

This calculation has the look of a `string miracle.'  It is not at all
obvious why the one-loop open string calculation should have given just this
result.  Had the R-R charges not satisfied the quantization condition, one
could likely use the argument from the first paragraph of this section to
show that the Type I theory has some sort of non-perturbative anomaly. 
A topological derivation of the D-brane charge has been given.\cite{ghs}

\subsection{The $p$-$p'$ System}

Simple $T$-duality gives parallel D-branes all with the same dimension but
now we are considering more general configurations.  In this section
we consider two D-branes, each parallel to the coordinate axes.\footnote
{This has been extended to D-branes at angles by Berkooz, Douglas, and
Leigh.\cite{bdl} }
An open string can have both ends on the same D-brane or one on each.  The
$p-p$ and $p'-p'$ spectra are the same as before, but the $p-p'$ strings are
new.  Since we are taking the D-branes to be parallel to the coordinate axes,
there are four possible sets of boundary conditions for each spatial
coordinate
$X^i$ of the open string, namely NN (Neumann at both ends), DD, ND, and DN.
What really will matter is the number $\nu$ of ND plus DN coordinates.  A
$T$-duality can switch NN and DD, or ND and DN, but $\nu$ is invariant.
Of course $\nu$ is even because we only have $p$ even or $p$ odd in a given
theory.

The respective mode expansions are 
\bea
{\rm NN\colon}&& X^\mu(z,\zb) = x^\mu - i\ap p^\mu
\ln(z\zb) + i\sqrt{\ap\over2} \sum_{m\neq0} {\alpha_m^\mu\over
m}(z^{-m}+\zb^{-m}), \nonumber\\
{\rm DN, ND\colon}&& X^\mu(z,\zb) = i\sqrt{\ap\over2} \sum_{r \in\Z +1/2}
{\alpha_r^\mu\over r}(z^{-r}\pm\zb^{-r}),
\label{modexps}\\
{\rm DD\colon}&& X^\mu(z,\zb) = -i \frac{\delta X^\mu}{2\pi} \ln(z/\zb)
+ i\sqrt{\ap\over2} \sum_{m\neq0}
{\alpha_m^\mu\over m}(z^{-m}-\zb^{-m})\ . \nonumber
\eea
In particular, the DN and ND coordinates have half-integer moding.
The fermions have the same
moding in the Ramond sector and opposite in the Neveu-Schwarz sector.  The
string zero point energy is 0 in the R sector as always, and
\be
(8-\nu)\left(-{1\over 24} - {1\over 48}\right) 
+ \nu \left({1\over 48} + {1\over 24}\right) = -{1\over 2} + {\nu\over 8}
\label{nszpe}
\ee
in the NS sector.

The oscillators can raise the level in half-integer
units, so only for $\nu$ a multiple of 4 is degeneracy between the R and NS
sectors possible.  Indeed, it is in this case that the $p$-$p'$ system is
supersymmetric.  Let us see this directly.  
As discussed in sections~3.2 and~3.3, a D-brane leaves unbroken the
supersymmetries
\be
Q_\alpha+ P {\tilde Q}_\alpha\ , \label{unb1}
\ee
where $P$ acts as a reflection in the direction transverse to
the D-brane.  With a second D-brane, the only unbroken supersymmetries will
be those that are also of the form
\be
Q_\alpha+ P' {\tilde Q}_\alpha = Q_\alpha+ P (P^{-1}P') {\tilde Q}_\alpha\ .
\ee
with $P'$ the reflection transverse to
the second D-brane.
Thus the unbroken supersymmetries correspond to the $+1$ eigenvalues of
$P^{-1}P'$.  In DD and NN directions this is trivial, while in DN and ND
directions it is a net parity transformation.  Since the number $\nu$ of
such dimensions is even, we can pair them and write $P^{-1}P'$ as a product
of rotations by $\pi$, 
\be
e^{i\pi (J_1 + \ldots + J_{\nu/2}) }\ .
\ee
In a spinor representation, each $e^{i\pi J}$ has eigenvalues $\pm
{i}$, so there will be unbroken supersymmetry only if $\nu$ is a
multiple of 4 as found above.\footnote{We will see that
there are supersymmetric {\it bound states} when $\nu = 2$.}

For example, 
Type~I theory, besides the 9-branes, will have 1-branes and 5-branes.
This is consistent with the fact that the only R-R field strengths are the
three-form and its Hodge-dual seven-form.  The 5-brane is required to have
two Chan-Paton degrees of freedom (which can be thought of as images under
$\Omega$) and so an $SU(2)$ gauge group.\cite{witinst}$^{\!-\,}$\cite{GP}

When $\nu = 0$, $P^{-1}P' = 1$ identically and there is a full
ten-dimensional spinor of supersymmetries.  This is the same as for the
original Type~I theory, to which it is $T$-dual.  In $D=4$ units, this is
$N=4$.  For $\nu=4$ or $\nu=8$ there is $D=4$ $N=2$ supersymmetry.

Let us now study the spectrum for $\nu = 4$, saving $\nu = 8$ for the next
lecture.  The NS zero-point energy is zero.  There are four periodic
world-sheet fermions $\psi^i$, namely those in the ND directions.  The four
zero modes generate $2^{4/2}$ or four ground states, of which two survive the
GSO projection.  In the R sector the zero-point energy is also zero; there are
four periodic transverse $\psi$, from the NN and DD directions not counting
$\mu=0,1$.  Again these generate four ground states of which two survive the
GSO projection.  The full content then is half of an $N=2$ hypermultiplet.
The other half comes from the world-sheet-orientation-reversed states: these
are distinct because for $\nu \neq 0$ the ends are always on different
D-branes.

Let us write the of the action for the bosonic $p-p'$ fields $\chi^A$,
starting with $(p,p') = (9,5)$.
Here $A$ is a doublet index under the $SU(2)_R$ of the $N=2$ algebra.  The
field $\chi^A$ has charges $(+1,-1)$ under the $U(1) \times U(1)$ gauge
theories on the branes.  The minimally coupled action
is then
\be
\int d^6\xi\, \left( \sum_{a=0}^5 \left| (\partial_a + iA_a - i A'_a) \chi
\right|^2 + 
\biggr(\frac{1}{4g_p^2} + \frac{1}{4g_{p'}^2} \biggl)
\sum_{I=1}^3 (\chi^{\dagger}\tau^I\chi)^2 \right)\ ,
\label{59act}
\ee
with $A_a$ and $A_a'$ the brane gauge fields, $g_p$ and $g_{p'}$ the
effective Yang-Mills couplings, and $\tau^I$ the $SU(2)$ matrices.  The second
term is from the
$N=2$ D-terms for the two gauge fields.  The integral is over the 5-brane
world-volume, which lies in the 9-brane world-volume.  Under $T$-dualities in
any of the ND directions, one obtains
$(p,p') = (8,6)$, $(7,7)$, $(6,8)$, or $(5,9)$, but the intersection of the
branes remains $(5+1)$-dimensional and the $p$-$p'$ strings live on the
intersection with action~(\ref{59act}).  In the present case the $D$-term is
nonvanishing only for $\chi^A = 0$, though more generally (say when there are
several coincident $p$ and $p'$-branes), there will be additional massless
charged fields and flat directions arise.

Under $T$-dualities in $r$ NN directions, one obtains $(p,p') = (9-r,5-r)$. 
The action becomes 
\bea
&&\int d^{6-r}\xi\, \left( \sum_{a=0}^{5-r} \left| (\partial_a + iA_a - i
A'_a) \chi
\right|^2\ + \frac{ \chi^{\dagger}\chi }{(2\pi\apm)^2 }
\sum_{a=6-r}^{5} ( X_a - X'_a )^2 \right.\nonumber\\
&&\qquad\qquad\qquad\qquad\qquad+\ 
\left.\biggl(\frac{1}{4g_p^2} + \frac{1}{4g_{p'}^2} \biggr)
\sum_{i=1}^3 (\chi^{\dagger}\tau^I\chi)^2 \right)\ .
\label{xyact}
\eea
The second term, proportional to the separation of the branes, is from the
tension of the stretched string.

\subsection{The BPS Bound}

The $N=2$ supersymmetry algebra (in a
Majorana basis) is
\bea
&& \{ Q_\alpha, Q_\beta \}\ 
=\ 2(\Gamma^0 \Gamma^\mu)_{\alpha\beta} ( P_\mu + Q^N_\mu/2\pi\ap )
\nonumber\\ 
&& \{ \tilde Q_\alpha, \tilde Q_\beta\}\ =\ 2(\Gamma^0
\Gamma^\mu)_{\alpha\beta} ( P_\mu - Q^N_\mu/2\pi\ap )
\nonumber\\ 
&& \{ Q_\alpha, \tilde Q_\beta \}\ =\ 2 \sum_{\{m\}} \tau_p (\Gamma^0
\Gamma^{m_1} \ldots \Gamma^{m_p})_{\alpha\beta} Q^R_{m_1\ldots m_p} \ .
\eea
Here $Q^N$ is the charge to which the NS-NS two-form couples and $Q^R$ are
the R-R charges, all normalized to one per unit world-volume.  The sum runs
over all ordered sets $m_1 < m_2 < \ldots < m_p$ such that
$p$ is even for IIA or odd for
IIB.  The R-R charges appear in the product of the right- and left-moving
supersymmetries, since the corresponding vertex operators are a product of
spin fields, while the NS-NS charges appear in right-right and left-left.

It is natural to define the dimensionless string coupling $g = e^\phi$ to be
the ratio of the fundamental (F-)string and D-string tensions in the IIB
theory, so that
\be
\tau_p = \frac{(2\pi\sqrt{\ap})^{1-p}}{2\pi\apm g}\ .
\ee
Comparing this with the string calculation~(\ref{dcharge}) fixes the relation
between $g$ and $\kappa = g \kappa_0$ and so determines the normalization
$\kappa_0$ of the spacetime action~(\ref{stact}),\cite{shanta}
\be
\kappa_0 = 8 \pi^{7/2} \ap^2\ .  \label{kap0}
\ee

As an example consider an object with the charges of $q_1$ F-strings
and $q_2$ D-strings in the IIB theory, at rest and aligned along the
one-direction.  The anticommutator implies
\be
\left\{ \left[ \begin{array}{c} Q_\alpha \\ \tilde Q_\alpha \end{array}
\right] , \left[ Q_\beta\ \tilde Q_\beta \right] \right\}
= \left[ \begin{array}{cc} 1&0 \\ 0&1 \end{array} \right] M +
\left[ \begin{array}{cc} q_1&q_2/g \\ q_2/g&-q_1 \end{array} \right]
\frac{v_1 \Gamma^0 \Gamma^1}{2\pi\apm}
\ee
where $v_1$ is the length of the system. 
The eigenvalues of $\Gamma^0 \Gamma^1$ are $\pm 1$ so those of the
right-hand side are $M \pm v_1 (q_1^2 + q_2^2/g^2)^{1/2}/2\pi\ap$.  The left
side is a positive matrix, giving the BPS bound on the tension\,\cite{schwarz}
\be
\frac{M}{v_1} \geq \frac{\sqrt{q_1^2 + q_2^2/g^2}}{2\pi\apm}\ . \label{fdbps}
\ee
This is saturated by the fundamental string, $(q_1,q_2) = (1,0)$, and by the
D-string, $(q_1,q_2) = (0,1)$. 

We leave it to the reader to extend this to a system with the quantum numbers
of Dirichlet $p$ and $p'$ branes. The result for $\nu$ a multiple of~4 is
\be
M \geq \tau_p v_p + \tau_{p'} v_{p'} \label{marg}
\ee
and for $\nu$ even but not a multiple of~4 it is\,\footnote{The difference
between the two cases comes from the relative sign of
$\Gamma^M (\Gamma^{M'})^T$ and $\Gamma^{M'} (\Gamma^{M})^T$. }
\be
M \geq \sqrt{\tau^2_p v^2_p + \tau^2_{p'} v^2_{p'} }\ .
\label{deep}
\ee
The branes are wrapped
on tori of volumes $v_p$ and $v'_p$ in order to make the masses finite.

The results~(\ref{marg}) and (\ref{deep}) are consistent with the earlier
results on supersymmetry breaking.  For $\nu$ a multiple of 4, a
separated $p$-brane and $p'$-brane do indeed saturate the
bound~(\ref{marg}).  For $\nu$ not a multiple of four, they do not saturate
the bound~(\ref{deep}) and cannot be supersymmetric.

\subsection{FD Bound States}

Consider a parallel D-string and F-string.  The total
tension
\be
\tau_{D1} + \tau_{F1} = \frac{g^{-1} + 1}{ 2\pi\apm}
\ee 
exceeds the BPS
bound~(\ref{fdbps}) and so this configuration is not supersymmetric.  However,
it can lower its energy as shown in figure~11.  (The whole discussion in
this section is based on Witten.\cite{witbound})
\begin{figure}
\begin{center}
\leavevmode
\epsfbox{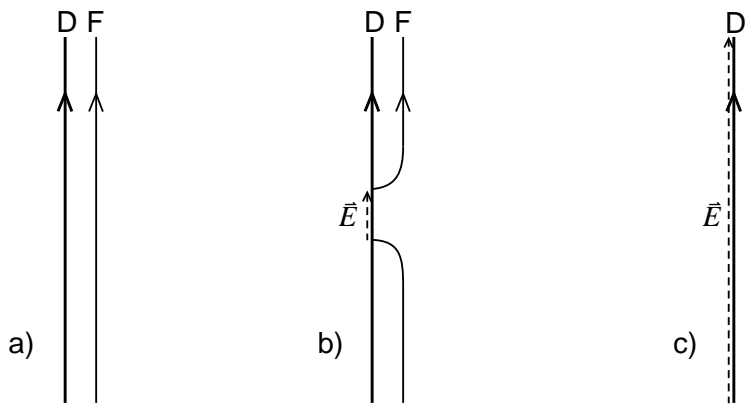}
\end{center}
\caption[]{a) Parallel D-string and F-string.  b) The F-string breaks, its
ends attaching to the D-string.  c) Final state: D-string with flux.}
\end{figure}
The F-string breaks, its endpoints attached to the D-string.  The endpoints
can then move off to infinity, leaving only the D-string behind.  Of course,
the D-string must now carry the charge of the F-string as well.  This comes
about because the F-string endpoints are charged under the D-string gauge
field, so a flux runs between them; this flux remains at the end.  Varying the
NS-NS $B_{\mu\nu}$ field in the D-brane action~(\ref{dact}), one sees that it
has a source proportional to the invariant flux $F_{ab} + B_{ab} / 2\pi\ap$. 
Thus the final D-string carries both the NS-NS and R-R two-form charges.  The
flux is of order $g$, its energy density is of order~$g$, and
so the final tension is $(g^{-1} + O(g))/2\pi\apm$.  This is below the tension
of the separated strings and of the same form as the BPS bound~(\ref{fdbps})
for a $(1,1)$ string.  A more detailed calculation shows that the final
tension saturates the bound,\cite{rract} so the state is supersymmetric.  In
effect, the F-string dissolves in the D-string, leaving flux behind.

To calculate the number of BPS states we should put the strings in a box of
length $L$ to make the spectrum discrete.  For the $(1,0)$ F-string, the usual
quantization of the ground state gives eight bosonic and eight fermionic
states moving in each direction for $16^2 = 256$ in all.  This is the
ultrashort representation of supersymmetry: half the 32 generators
annihilate the BPS state and the other half generate $2^8 = 256$ states.  The
same is true of the $(0,1)$ D-string and the $(1,1)$ bound state just found,
as will be clear from the discussion of the D-string in the next lecture.

Incidentally, the $(1,0)$ F-string leaves unbroken half the supersymmetry and
the $(0,1)$ D-string leaves unbroken a different half of the supersymmetry. 
The $(1,1)$ bound state leaves unbroken not the intersection of the two
(which is empty), but yet a different half.  The unbroken symmetries are
linear combinations of the unbroken and broken supersymmetries of the
D-string.

All the above extends immediately to $q$ F-strings and one D-string, forming
a supersymmetric $(q,1)$ bound state.  The more general case of $q_1$
F-strings and $q_2$ D-strings is more complicated.  The gauge dynamics are
now non-Abelian, the interactions are strong in the infrared, and no explicit
solution is known.  When $q_1$ and $q_2$ have a common factor, the BPS bound
makes any bound state only neutrally stable against falling apart into
subsystems.  To avoid this complication let $q_1$ and $q_2$ be relatively
prime, so any supersymmetric state is discretely below the continuum of
separated states.  This allows the Hamiltonian to be deformed to a simpler
supersymmetric Hamiltonian whose supersymmetric states can be 
determined explicitly, and again there is one ultrashort
representation, $256$ states.  The details, which are a bit intricate,
are left to the original reference.\cite{witbound}

\subsection{$0$-$p$ Bound States}

Bound states of $p$-branes and $p'$-branes have many applications.  Here we
focus on $p' = 0$, which can in general be reached by a $T$-duality.

{\it 0-0 bound states:} 

The BPS bound for the quantum numbers of $n$ 0-branes
is $n \tau_0$, so any bound state will be at the edge of the continuum.  To
make the bound state counting well defined, compactify one direction and give
the system momentum $m/R$ with $m$ and $n$ relatively prime.\cite{senbound} 
The bound state now lies discretely below the continuum, because the momentum
cannot be shared evenly among unbound subsystems.

This bound state problem is $T$-dual to the one just considered.  Taking the
$T$-dual, the $n$ 0-branes become D 1-branes, while the momentum becomes
winding number, corresponding to $m$ F-strings.  There is therefore one
ultrashort multiplet of supersymmetric states when $m$ and $n$ are relatively
prime.\cite{senbound} This bound state should still be present back in infinite
volume, since one can take $R$ to be large compared to the size of the bound
state.

{\it 0-2 bound states:}\,\footnote
{This section is based on conversations with 
J. Harvey, G. Moore, and A. Strominger.}

Now the BPS bound~(\ref{deep}) puts any bound
state discretely below the continuum.  One can see a hint of a bound state
forming by noticing that for a coincident 0-brane and 2-brane the NS 0-2
string has a negative zero-point energy~(\ref{nszpe}) and so a tachyon
(which survives the GSO projection), indicating instability towards
something.  In fact the bound state (one short representation) is easily
described: the 0-brane dissolves in the 2-brane, leaving  flux.  The brane R-R
action~(\ref{csact}) contains the coupling
$C_{(1)} F$, so with the flux the 2-brane also carries the 0-brane
charge.\cite{towndf} There is also one short multiplet for $n$ 0-branes.  This
same bound state is always present when $\nu = 2$.

{\it 0-4 bound states:} 

The BPS bound~(\ref{marg}) makes any bound state
marginally stable, so the problem is made well-defined as in the 0-0 case by
compactifying and adding momentum.\cite{senbound2}  The interactions in the
action~(\ref{xyact}) are relevant in the infrared so this is again a hard
problem, but as before it can be deformed into a solvable supersymmetric
system.  Again there is one multiplet of bound states.\cite{senbound2}
Now, though, the bound state is invariant only under $\frac{1}{4}$ of the
original supersymmetry, the intersection of the supersymmetries of the
0-brane and of the 4-brane.  The bound states then lie in a short (but not
ultrashort) multiplet of $2^{12}$ states.

For $2$ 0-branes and one 4-brane, one gets the correct count as
follows.\cite{vafa1} Think of the case that the volume of the 4-brane is
large.  The $16$ supersymmetries broken by the 4-brane generate $256$ states
that are  delocalized on the 4-brane.  The 8 supersymmetries unbroken by the
4-brane  and broken by the 0-brane generate $16$ states (half bosonic and
half fermionic), localized on the 0-brane. The total number is the product
$2^{12}$.  Now count the number of ways two 0-branes can be put into their
$16$ states on the 4-brane: there are $8$ states with both 0-branes in the
same (bosonic) state and $\ha 16\cdot 15$ states with the D-branes in
different states, for a total of $8\cdot 16$ states.  But in addition, the
two-branes can bind, and there are again $16$ states where the bound state
binds to the 4-brane.  Indirect arguments for the existence of these bound
states have been given,\cite{vafa1} and presumably it
can be demonstrated directly by the
method used for 0-0 bound states in
free space.  The total,
tensoring again with the 4-brane ground states, is $9 \cdot 16 \cdot 256$.

For $n$ 0-branes and one 4-brane, the degeneracy $D_n$ is given by the
generating functional\,\cite{vafa1}
\be
\sum_{n=0}^\infty q^n D_n\ =\ 256\ \prod_{k=1}^\infty \left( \frac{1 + q^k}{1
- q^k}\right)^8\ ,  \label{degen}
\ee
where the term $k$ in the product comes from bound states of $k$ 0-branes
then bound to the 4-brane.

{\it 0-6 bound states:} 

The relevant bound is~(\ref{deep}) and again any
bound state would be below the continuum.  The NS zero-point energy for 0-6
strings is positive, so there is no sign of decay.  One can give 0-brane
charge to the 6-brane
by turning on flux, but there is no way to do this and saturate the
BPS bound.  So it appears that there are {\it no} supersymmetric bound
states.  Incidentally, and unlike the 0-2 case, the 0-6 interaction is
repulsive, both at short distance and\,\cite{horpriv} at long.

{\it 0-8 bound states:} 

I am confused by this case.  There are two recent
papers with some relevant observations.\cite{08}

\section{Lecture 5: Applications}

\subsection{String Duality}

Now we can use D-branes to understand the strong-coupling limits of the
theories in which they exist.  

{\it Type IIB:}

Consider the D 1-brane of the IIB theory. The gauge field has no local
dynamics, so the only bosonic excitations are the transverse fluctuations.
The GSO projection on the open string Ramond ground state correlates the
$SO(8)$ chirality with the right/left motion, so the right-moving spinors on
the D-string are in the $\bf 8_s$ of $SO(8)$, and the left-moving spinors in
the $\bf 8_c$.
These are the same as the fluctuations of a
fundamental IIB string.\cite{witbound}  There, the supersymmetries
$Q_\alpha$ and $\tilde Q_\alpha$ have the same chirality.  Half of each
spinor annihilates the F-string and the other half generates fluctuations.
Since the supersymmetries have the same $SO(9,1)$ chirality, the $SO(8)$
chirality is correlated with the direction of motion.

Up until now we have used the string metric, but for discussing string
duality it is useful to switch to the Einstein metric, $g_{\mu\nu}^{\rm (E)}
= g^{-1/2} g_{\mu\nu}^{\rm (s)}$.   The Einstein metric, for which the
gravitational action has no dependence on the dilaton, is an invariant under
duality.
The tensions are then
\bea
\mbox{F-string:} && g^{1/2} / 2\pi\apm \nonumber\\
\mbox{D-string:} && g^{-1/2} / 2\pi\apm\ .
\eea
These are exact because of the
BPS property.  

At weak coupling the D-string is heavy and the F-string
tension is the lightest scale in the theory.  At strong coupling, however,
the D-string is the lightest object in the theory,\footnote
{In particular, a dimensional argument shows that the lowest-dimensional
branes have the lowest scale.\cite{hullscale}} and it is natural to believe
that the theory can be reinterpreted as a theory of weakly coupled
D-strings, with
$g' = g^{-1}$.  One cannot prove this without a nonperturbative definition
of the theory, but quantizing the light D-string implies a large number of
the states that would be found in the dual theory, and (Occam's
razor) self-duality of the IIB theory seems by far the simplest
interpretation---given that physics below the Planck
energy is described by some specific string theory, it seems likely that
there is a unique extension to higher energies.  This agrees with the duality
deduced from the low energy action and other
considerations.\cite{hullt}$^{\!-\,}$\cite{wit} In particular, the NS-NS and
R-R two-form potentials, to which the D- and F-strings respectively couple,
are interchanged by this duality.\footnote
{This type of argument had been applied to NS solitons in the context
of six-dimensional string-string duality.\cite{hetsol}}

The full duality group of the $D=10$ Type IIB theory is expected to be
$SL(2,Z)$.\cite{hullt}$^{\!-\,}$\cite{wit}  This relates the fundamental
string not only to the R-R string but to a whole set of strings with the
quantum numbers of $q_1$ F-strings and $q_2$ D-strings for $q_1$ and $q_2$ 
relatively prime.\cite{schwarz}  The bound states found in section~4.4 are
just what is required for $SL(2,Z)$ duality.\cite{witbound}  As the
coupling and the R-R scalar are varied, each of these strings becomes light
at the appropriate point in moduli space.

{\it Type I:}

For the D-strings of the Type I theory there are two
modifications.\cite{polwit}
The first is the projection onto oriented states.  The $U(1)$ gauge field,
with vertex operator $\partial_t X^\mu$, is removed.  The collective
coordinates, with vertex operators
$\partial_n X^\mu$, remain in the spectrum because the normal derivative is
even under $\Omega$.  That is, in terms of its action on the $X$ oscillators
$\Omega$ has an additional $-1$ for the $m=2,\ldots,9$
directions, as compared to the action on the usual NN strings.  By
superconformal symmetry this must extend to the fermions, so that on the
ground states $\Omega$ is no longer the identity but acts as
$R=e^{i\pi(S_1+S_2+S_3+S_4)}$.  This removes the left-moving $\bf 8_c$ and
leaves the right-moving $\bf 8_s$ (or vice versa: we have made an arbitrary
choice in defining $R$).

The second modification is the inclusion of 1-9 strings, strings with
one end on the 1-brane and one on a 9-brane, the latter corresponding
to the usual $SO(32)$ Chan-Paton factor.  This is the case $\nu = 8$ from
section~4.2.
\begin{figure}
\begin{center}
\leavevmode
\epsfbox{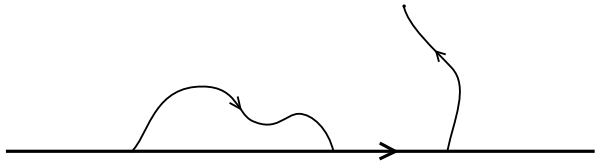}
\end{center}
\caption[]{D-string in Type I theory with attached 1-1 and 1-9 strings.}
\end{figure}
The $\Omega$
projection determines the 9-1 state in terms of the 1-9, but
otherwise makes no constraint.  The zero point energy~(\ref{nszpe}) is
strictly positive so there are no massless states in the NS
sector.  The R ground states are, as always, massless.  There are two ground
states, from the periodic
$\psi^{0,1}$ oscillators, 
\be
|\pm; i\rangle = ( \psi^0_0 \pm \psi^1_0 ) |i\rangle,
\ee 
where $i$ is the Chan-Paton index for the 9-brane end.  One of the two
states
$|\pm; i\rangle$ is removed by the GSO projection, and the $G_0$ physical
state condition on the remaining state implies that these massless fermions
are chiral on the one-brane.  Spacetime supersymmetry can only be
satisfied if they move oppositely to the 1-1 fermions; at the world-sheet
level this would have to follow from a careful analysis of the OPE of the
gravitino. The 1-9 strings, having one 9-brane
Chan-Paton index, are vectors of $SO(32)$.

Thus the world-sheet theory of the Type I one-brane is precisely that of
the $SO(32)$ heterotic string, 
with the spacetime supersymmetry realized in Green-Schwarz form and the
current algebra in fermionic form.\cite{polwit}  As in the IIB theory we can
now argue that this $SO(32)$ heterotic string sets the lightest scale in the
theory when $g$ becomes large, and so the natural assumption is that the
strongly coupled behavior is governed by the weakly coupled heterotic
string.  This is in agreement with the conclusion from various other
arguments.\cite{wit}$^{\!,\,}$\cite{blackone}

The fermionic $SO(32)$ current algebra requires a GSO projection.  It is
interesting to see how this arises in the D-string.  Consider a closed
D-string.  The $\Omega$ projection removed the $U(1)$ gauge field, but is
consistent with a discrete gauge symmetry, a holonomy $\pm 1$ around the
D-brane.  This discrete gauge symmetry is the GSO projection, and evidently
the rules of D-branes require us to sum over all consistent possibilities in
this way.
We can now see how D-strings account for the spinor
representation of $SO(32)$ in the Type~I theory.  In the {R} sector of
the discrete D-brane gauge theory, the
1-9 strings are periodic.  The zero modes of the fields $\Psi^i$,
representing the massless
1-9 strings, satisfy the Clifford algebra
\be
\{ \Psi^i_0 , \Psi^j_0 \} = \delta^{ij}, \qquad i,j= 1,
\cdots , 32 .
\ee
The quantization now proceeds just as for the fundamental heterotic string,
giving spinors ${\bf 2^{31}} + \overline{\bf  2^{31}}$, one of which is
removed by the discrete gauge symmetry.

\subsection{The Eleventh Dimension}

One of the greatest surprises in string duality was the
discovery of a new dimension, the eleventh.  Let us see how we can find this
dimension with the aid of D-branes. 

\indent{\it Type IIA:} 

In the IIA theory, the 0-brane has a mass
$\tau_0 = \ap^{-1/2}g$ in the string metric.  As $g \to \infty$, this
mass is the lightest scale in the theory.  In addition, we have seen in the
previous lecture that $n$
0-branes have a single supersymmetric bound state with mass $n\tau_0$. 
This evenly spaced tower of states is characteristic of the appearance of an
additional dimension, where the momentum (Kaluza-Klein) states have masses
$n/R$ and form a continuum is $R \to \infty$.  The existence of these
states and of the eleventh dimension was inferred even before the significance
of D-branes was understood, because they are required by lower-dimensional
$U$-dualities.\cite{town}$^{\!-\,}$\cite{wit}$^{\!,\,}$\footnote
{In contrast, the interpretation of the conifold singularity in terms of a
massless soliton requires that there not be bound states.  This is a
different problem, because the D-brane is now wrapped on a curved space,
and the required spectrum has been found.\cite{bsv}} 

To relate the coupling to the size of the eleventh dimension we need to
compare the respective actions,\cite{wit}
\be
\frac{1}{2\kappa_0^2 g^2} \int d^{10} x\, \sqrt{-g_{\rm s}} R_{\rm s}
= \frac{2\pi R}{2\kappa_{11}^2} \int d^{10} x\, \sqrt{-g_{11}} R_{11}\ .
\ee
The string and M theory metrics are equal up to a rescaling,
\be
g_{{\rm s}\mu\nu} = \zeta^2 g_{{\rm 11}\mu\nu}  \label{mmet}
\ee
and so $\zeta^8 = 2\pi R \kappa_0^2 g^2 / \kappa_{11}^2$.
The respective masses are related $nR^{-1} = m_{11}
= \zeta m_{\rm s} = n \zeta \tau_0$ or
$R = \ap^{1/2} g/\zeta$.  Combining these with the result~(\ref{kap0}) for
$\kappa_0$, we obtain 
\be
\zeta = g^{1/3} \left[2^{7/9} \pi^{8/9} \ap \kappa_{11}^{-2/9}
\right]
\ee
and 
\be
R = g^{2/3} \left[2^{-7/9} \pi^{-8/9} \kappa_{11}^{2/9} \right]\ .  \label{mrad}
\ee
In order to emphasize the basic structure we
hide in braces numerical factors and factors of
$\kappa_{11}$ and $\ap$.  The former we hope are correct;\,\footnote
{See also.\cite{shanta}} the
latter are determined by dimensional analysis, with $\kappa_{11}$ having
units of (M theory length$^{9/2}$) and $\ap$ (string theory length$^2$). 
We are free to set $\zeta = 1$, using the same metric and units in M theory as
in string theory.  In this case
\be
\kappa_{11}^2 = g^3 \left[2^7 \pi^8 \ap^{9/2} \right]\ .
\ee
The reason for not always doing so is that when we have a series of dualities,
as below, there will be different string metrics.

It is interesting to track the eleven-dimensional origin of the various
branes of the IIA theory.  The D 0-branes are, as we have just seen,
Kaluza-Klein states. The F 1-branes, the IIA strings themselves, are
wrapped membranes of M-theory.\cite{supmem}  The D 2-branes are membranes
transverse to the eleventh dimension $X^{10}$.  The D 4-branes are M-theory
5-branes wrapped on $X^{10}$, while the NS (symmetric) 5-branes are
M-theory 5-branes transverse to $X^{10}$.  The D 6-branes, being the
magnetic duals of the 0-branes, are Kaluza-Klein monopoles.  The
D 8-branes are problematic, as there would seem to be no M-theory 8- or
9-branes from which they can descend.  The point is that the D 8-branes,
being sources of co-dimension one, cause the dilaton to diverge within a
finite distance.\cite{polwit}  They must therefore be a finite distance from
an orientifold plane, which is essentially a boundary of spacetime.  As
the coupling grows, the distance to the divergence and the boundary
necessarily shrinks, so that they disappear into it in the strong coupling
limit: they become part of the gauge dynamics of the nine-dimensional
boundary of M-theory,\cite{horwit} to be discussed in more detail below.

One can see further indication of the eleventh dimension in the dynamics
of the D 2-brane.  In $2+1$ dimensions, the vector field on the brane is
dual to a scalar, through Hodge duality of the field strength, $*F_2 =
d\phi$.  This scalar is the eleventh embedding 
dimension.\cite{duff11}$^{\!,\,}$\cite{towndf}
Carrying out the duality in detail, the D 2-brane action is found to have
a hidden eleven-dimensional Lorentz invariance!

{\it $E_8 \times E_8$ Heterotic String:}

We have deduced the duals of four of the five ten-dimensional string
theories.  Let us study the final one, the $E_8 \times E_8$ heterotic
string, by using its $T$-duality to the $SO(32)$ string.\cite{narain}
Compactify on a large radius $R$ and turn on a Wilson line which breaks
$E_8\times E_8$ to $SO(16) \times SO(16)$.  This is $T$-dual to the $SO(32)$
heterotic string, again with a Wilson line breaking the group to 
$SO(16) \times SO(16)$.  The couplings and radii are related
\bea
R' &=& R^{-1} \left[ \apm \right] , \nonumber\\
g' &=& g R^{-1} \left[ {\ap}^{1/2} \right] \ .
\eea
Now use Type~I - heterotic duality to write this as a Type~I theory
with\,\cite{wit}
\bea
R_{\rm I} &=& g'^{-1/2} R'\ =\ g^{-1/2} R^{-1/2} \left[ \apm^{3/4} \right],
\nonumber\\ 
g_{\rm I} &=& g'^{-1}\ =\ g^{-1} R \left[ {\ap}^{-1/2} \right]\ .
\eea
The radius is very small, so it is useful to make another $T$-duality, to
the `Type I$'$' theory.  The compact dimension is then a segment
of length $\pi R_{\rm I'}$ with
eight D 8-branes at each end, and
\bea
R_{\rm I'} &=& R_{\rm I}^{-1} \left[ \apm \right] \ =\ g^{1/2}
R^{1/2} \left[  \apm^{1/4} \right], 
\nonumber\\
g_{\rm I'} &=& g_{\rm I} R_{\rm I}^{-1} \left[ 2^{-1/2} {\ap}^{1/2} \right]\ =\
g^{-1/2} R^{3/2} \left[ 2^{-1/2} \apm^{-3/4} \right]\ .
\eea
Now take $R \to \infty$ to recover the original ten-dimensional theory
(in particular the Wilson line is irrelevant and the original $E_8\times
E_8$ restored).  Both the radius and the coupling of the Type I$'$ theory
become large.  The physics between the ends of the segment is given locally
by the IIA string, and so the strongly coupled limit is
eleven-dimensional.  Taking into account the transformations~(\ref{mmet}),
(\ref{mrad}), the radii of the two compact dimensions in M-theory units are
\bea
R_9 &=& \zeta_{\rm I'}^{-1} R_{\rm I'}\ =\ 
g^{2/3} \left[2^{-11/18} \pi^{-8/9} \kappa_{11}^{2/9} \right]\label{r910}
\\ 
R_{10} &=&
g_{\rm I'}^{2/3} \left[2^{-7/9} \pi^{-8/9} \kappa_{11}^{2/9} \right]
\ =\ 
g^{-1/3} R \left[2^{-10/9} \pi^{-8/9} \ap^{-1/2} \kappa_{11}^{2/9} \right]\ .  
\nonumber
\eea
As $R\to \infty$, $R_{10} \to\infty$ also, while $R_9$ remains fixed and
(for $g$ large) large compared to the Planck scale.  Thus, in the strongly
coupled limit of the ten-dimensional $E_8\times E_8$ heterotic string an
eleventh dimension again appears, a segment of length $R_9$,
with one $E_8$ factor on each endpoint.\cite{horwit} 

\subsection{$U$-Duality}

An interesting feature of string duality is the enlargement of the duality
group under further toroidal compactification.  Understanding the origin of
these $U$-dualities, perhaps in some geometric way, might be a useful avenue
to finding the underlying structure in string theory.  Here I would like to
describe one example, the Type~II string on a five-torus.  This is chosen
because it is the setting for the simplest black hole state counting, and
also because the necessary group theory is somewhat familiar from grand
unification.

Let us first count the gauge fields.  From the NS-NS sector there are 5
Kaluza-Klein gauge bosons and 5 gauge bosons from the antisymmetric tensor. 
There are 16 gauge bosons from the dimensional reduction of the various R-R
forms.  Finally, one can form a field strength from the Hodge dual $*H$ of
the 3-form field strength of the NS-NS $B_{\mu\nu}$.

Let us see how $T$-duality acts on these.  The $T$-duality is $SO(5,5;\Z)$,
generated by $T$-dualities on the various axes, linear redefinitions of the
axes, and discrete shifts of the antisymmetric tensor.  This mixes the first
10 NS-NS gauge fields among themselves, and the 16 R-R gauge fields among
themselves, and leaves the final NS-NS field invariant.
The $SO(5,5;\Z)$ representations here correspond directly to the {\bf 10},
{\bf 16}, and {\bf 1} of $SO(10)$.

The low energy supergravity theory for this compactification has a
continuous symmetry $E_{6(6)}$ which is a noncompact version of
$E_6$.\cite{julia}  This is one of those supergravity properties that was
ignored for some time, because there is no sign of it in (perturbative)
string theory. But now we know better:\,\cite{hullt} a discrete subgroup
$E_{6(6)}(\Z)$ is supposed to be a good symmetry of the full theory.

The gauge bosons are in the {\bf 27} of $E_{6(6)}(\Z)$, which is the same as
the {\bf 27} of $E_{6(6)}$.  The decomposition under $SO(10)
\sim SO(5,5;\Z)$ is familiar from grand unified model building,
\be
{\bf 27}\ \to\ {\bf 10}\ +\ {\bf 16}\ +\ {\bf 1}\ .
\ee
The excitations carrying the {\bf 10} charges are just the Kaluza-Klein and
winding strings.  The $U$-duality requires also states in the {\bf 16}.
These are just the various wrapped D-branes.  Finally, the state
carrying the {\bf 1} charge is an ordinary NS-NS soliton, a symmetric
5-brane.\cite{chs}

{\it $U$-Duality and Bound States:}

It is interesting to see how some of the bound state results from the 
previous lecture fit the predictions of $U$-duality.  We will generate
$U$-transformations as a combination of $T_{mn\cdots p}$, which is a
$T$-duality in the indicated directions, and $S$, the IIB weak/strong
transformation.  The former switches between N and D boundary conditions and
between momentum and winding number in the indicated directions.  The latter
interchanges the NS and R two-forms but leaves the R four-form invariant, and
acts correspondingly on the solitons carrying these charges.  We denote by
$\rD_{mn\cdots p}$ a D-brane extended in the indicated directions, and
similarly for $\rF_m$ a fundamental string and $p_m$ a momentum-carrying BPS
state.

The first duality chain is
\be
(\rD_9,\rF_9)\ \stackrel{T_{78}}{\to}\ (\rD_{789},\rF_9)
\ \stackrel{S}{\to}\ (\rD_{789},\rD_9)
\ \stackrel{T_{9}}{\to}\ (\rD_{78},\rD_{\emptyset})\ .
\ee
Thus the D-string--F-string bound state is $U$-dual to the 0-2 bound state.

The second chain is
\be
(\rD_{6789},\rD_{\emptyset})\ \stackrel{T_{6}}{\to}\ (\rD_{789},\rD_6)
\ \stackrel{S}{\to}\ (\rD_{789},\rF_6)
\ \stackrel{T_{6789}}{\to}\ (\rD_{6},p_6)
\ \stackrel{S}{\to}\ (\rF_{6},p_6)
\ee
The bound states of $n$ 0-branes and $m$ 4-branes are thus $U$-dual to
fundamental string states with momentum $n$ and winding number $m$.  The
bound state degeneracy~(\ref{degen}) for $m=1$ precisely matches the
fundamental string
degeneracy.\cite{vafwit}$^{\!,\,}$\cite{senbound2}$^{\!-\,}$\cite{vafa2}
For $m>1$ the same form~(\ref{degen}) should hold but with $n \to mn$.  This
is believed to be the case, but the analysis (which requires the
instanton picture described in the next section) does not seem to be
complete.\cite{vafa2}

A related issue is the question of branes ending on other
branes.\cite{andyopen}  An F-string can of course end on a D-string, so from
the first duality chain it follows that a D $p$-brane can end on a D
$(p+2)$-brane.  The key issue is whether the coupling between
spacetime forms and world-brane fields allows the source to be conserved, as
with the NS-NS two-form source in figure~11.  Similar arguments can then be
applied to the extended objects in M
theory.\cite{andyopen}$^{\!,\,}$\cite{towndf}

\subsection{D-Branes as Instantons}

Consider now a 0-brane and $n$ coincident 4-branes.  The potential terms in
the action are
\be
\frac{ \chi_i^{\dagger} \chi_i }{(2\pi\apm)^2 }
\sum_{a=1}^{5} ( X_a - Y_a )^2 + \frac{1}{4g_0^2}
\sum_{I=1}^3 (\chi_i^{\dagger}\tau^I\chi_i)^2 \ .
\label{04pot}
\ee
Here $a$ runs over the dimensions transverse to the 4-brane, and $X_a$ and
$Y_a$ are respectively the 0-brane and 4-brane positions.   This is the same
as in the earlier action~(\ref{xyact}), except that we have taken the
4-branes to have infinite volume so that their
$D$-term drops out, and the 0-4 field $\chi$ now carries a 4-brane index
$i$ indicated explicitly (in addition to the $SU(2)_R$ index which is
suppressed).  The potential~(\ref{04pot}) is exact on grounds of $N=2$
supersymmetry.  The first term is the $N=2$ coupling between the
hypermultiplets $\chi$ and the vector multiplet scalars $X$, $Y$, while
the second is the $U(1)$ $D$-term.

For $N > 1$ there are two branches of moduli space,
\bea
{\rm Coulomb\ branch\colon}&& X \neq Y, \quad \chi = 0\nonumber\\
{\rm Higgs\ branch\colon}&& X = Y, \quad \chi \neq 0.
\eea
The Coulomb moduli space is just position space for the 0-brane.
On the Higgs branch on the other hand, the 0-brane has gotten stuck on the
4-branes.  In fact, it has dissolved into gauge fields, just as in the
bound state discussions from the previous lecture.  The R-R couplings
include a term $C_{(1)} {\rm Tr}(F^2)$, so that when there is an instanton on
the 4-brane it carries the 0-brane charge.  One can also check this from
the dimension of moduli space.  There are $4N$ real degrees of freedom in
$\chi$.  The vanishing of the $U(1)$ $D$-term imposes three constraints,
and modding by the (broken) $U(1)$ removes another degree of freedom
leaving $4N - 4$.  In addition there are the 4 moduli for the position of the
0-brane in the directions parallel to the 4-branes, for a total of $4N$
moduli.  This is the correct number for an $SU(N)$ instanton when we do not
mod out also the $SU(N)$ identifications.  For $k$ instantons this becomes
$4Nk$.

The connection between D-branes and instantons was found first in the case
$(p,p') = (9,5)$ by Witten.\cite{witinst}  This is $T$-dual to the above but
does not have the Coulomb branch.  That a $p$-brane instanton can shrink to
zero size and move off as a $(p-4)$-brane was noted by
Douglas.\cite{douginst}

Note that this result has nothing to do with string duality, since the
string coupling can be held fixed and small.  Rather, it is a conformal
field theory result.  Starting from large instantons, it has
always been a puzzle how to understand instanton amplitudes when the size
of the instanton drops below the string scale.  We now see, at least for
instantons coupled to boundary (Chan-Paton) gauge fields, that there is a
simple expansion around the zero-size limit.

\subsection{D-Branes as Probes}

In the preceding section we interpreted the 0-4 strings $\chi$ as moduli for the
size and orientation of an instanton obtained as a blown-up D-brane.  In a very
interesting paper, Douglas has shown that one can see the resulting instanton by
using another D-brane as a probe.\cite{dougprobe}  First take a $T$-dual so that
the gauge fields live on 9-branes and the instanton is a 5-brane; then add a
1-brane as a probe.  When the 1-brane is close to the 5-brane, closer than the
string scale, the dynamics is dominated by the light open strings, especially
the short 1-5 strings.  Integrating out the light open strings to find the
effective gauge field seen by the probe, one recovers Witten's lagrangian
version of the ADMH construction and so the
instanton gauge field.\cite{witadmh}$^{\!,\,}$\cite{admh}  In particular, the
variety of fields in that construction arise as the various
$p$-$p'$ strings of the D-brane system.

Thus it seems that one can sensibly discuss an instanton smaller than the
string scale.  This is in conflict with the idea that the
string length is a minimum distance.  Indeed, there were arguments to this
effect made by Shenker,\cite{shenk2} and there have been recent studies of
D-brane scattering suggesting that D-branes do indeed probe a shorter length
scale.\cite{bachas}$^{\!,\,}$\cite{short} This should be covered in detail in
Steve's lectures.  There have been many other interesting recent applications of
this idea, in particular by Seiberg, but I will not have time to pursue this.

I will sketch a different example of the D-brane as probe,\cite{tensors}
similar in nature to the instanton but simpler in its field content.  This is
the blowing up of a $\Z_2$ fixed point.  It is known that twisted-sector moduli
have the interpretation of smoothing out the orbifold geometry.  With a D-brane
probe this can be seen directly.

Consider the $\R^6 \times (\R^4/\Z_2)$ orbifold formed by the reflection $R$,
$X^{6,7,8,9} \to -X^{6,7,8,9}$, in the IIB theory.  We are focusing
on the neighborhood of a single fixed point.  Add a D-string in this plane at
$X^{2,\ldots,9} = 0$.  In order for the string to be able to move off the fixed
plane it needs two Chan-Paton indices, for the string and its $\Z_2$ image. 
Since $R$ takes the D-string into its image, it acts on an open string state as
\be
R |{\psi,ij}\rangle = \sigma^1_{ii'} |{R\psi,i'j'}\rangle \sigma^1_{j'j}\ .
\ee
That is, it acts on the oscillators in the usual way and also switches the
Chan-Paton factors for the brane and its image.  This is a simple example of
the general formalism for orbifolds and orientifolds with D-branes.\cite{GP}
In the NS sector, the massless $R$-invariant states are then
\begin{eqnarray}
&&\partial_t X^\mu \sigma^{0,1}, \qquad \mu = 0,1 \nonumber\\
&&\partial_n X^i \sigma^{0,1}, \qquad i = 2,3,4,5 \nonumber\\ 
&&\partial_n X^m \sigma^{2,3}, \qquad m = 6,7,8,9.
\end{eqnarray}
These are respectively a gauge field, the position of the string within
the six-plane, and the transverse position.  Call the corresponding
D-string fields $A^\mu, x^i, x^m$, all $2\times2$ matrices.  The
bosonic action is the $d = 10$ $U(2)$ Yang-Mills action, dimensionally
reduced and $R$-projected (which breaks the gauge symmetry to $U(1)
\times U(1))$.  In particular, the potential is
\be
U = 2\sum_{i,m} {\rm Tr}([x^i,x^m]^2) + \sum_{m,n} {\rm
Tr}([x^m,x^n]^2). \label{dppot}
\ee
The moduli space thus has two branches.  On one, $x^m = 0$ and $x^i =
u^i \sigma^0 + v^i \sigma^1$.  This corresponds to two D-strings moving
independently in the plane, with positions $u^i \pm v^i$.  The gauge
symmetry is unbroken, giving independent $U(1)$'s on each
D-string.
On the other branch, $x^m$ is nonzero and $x^i = u^i
\sigma^0$.  The $\sigma^1$ gauge invariance is broken and so by gauge
choice $x^m = w^m \sigma^3$.  This corresponds to the D-string moving
off the fixed plane, the string and its image being at
$(u^i, \pm w^m)$.

Now let us turn on twisted-sector moduli.  Define complex $q^m$ 
by $x^m = \sigma^3 {\rm Re}(q^m) + \sigma^2 {\rm Im}(q^m)$, and define two
doublets,
\begin{equation}
\Phi_0 = \left( \begin{array}{c} q^6 + i q^7 \\ q^8 + i q^9
\end{array} \right), \qquad
\Phi_1 = \left( \begin{array}{c} \bar q^6 + i \bar q^7 \\ \bar q^8 + i
\bar q^9
\end{array} \right).
\end{equation}
These have charges $\pm 1$ respectively under the $\sigma^1$ $U(1)$.
The three NSNS moduli can be written as a vector $\D$,
and the potential is proportional to
\begin{equation}
(\Phi_0^\dagger \mbox{\boldmath$\tau$} \Phi_0 - \Phi_1^\dagger
\mbox{\boldmath$\tau$} \Phi_1 + \D)^2,
\end{equation}
where the Pauli matrices are now denoted $\tau^I$ to emphasize that they
act in a different space.  This reduces to the second term of the
earlier potential~(\ref{04pot}) when $\D = 0$.  Its form
is determined by supersymmetry.

For $\D \neq 0$ the
orbifold point is blown up.  The moduli space
of the D-string is simply the set of possible locations, that is, the
blown up ALE space.\footnote
{Note that the branch of the
moduli space with $v^i \neq 0$ is no longer present.}  The $z^m$ contain
eight scalar fields.  Three are removed by the $\D$-flatness condition
that the potential vanish, and a fourth is a gauge degree of freedom,
leaving the expected four moduli.  In terms of supermultiplets, the
system has the equivalent of $d=6$ $N=1$ supersymmetry.  The D-string
has two hypermultiplets and two vector multiplets, which
are Higgsed down to one
hypermultiplet and one vector multiplet. 
The idea of
Douglas\,\cite{dougprobe} is that the metric on this moduli space, as seen in
the kinetic term for the D-string fields, should be the smoothed orbifold
metric.  It is straightforward to verify this, and we omit the
details.\cite{tensors}  The result is the Eguchi-Hanson metric,\cite{ale} which
is the correct (hyper-K\"ahler) metric on the blown-up orbifold.  This
construction is known as the hyper-K\"ahler quotient.\cite{hyper}  This
geometry seems to make sense even when its scale is far shorter than the
string scale.  This has recently been extended to other
orbifolds.\cite{cvjmyers}

\subsection{D-Branes as Black Holes}

Before the significance of D-branes was recognized, the R-R charged objects
required by string duality were believed to be black
$p$-branes.\cite{hullt}$^{\!-\,}$\cite{coni}
Such solutions exist for any tensor gauge charge,\cite{blackp} with the source
essentially hidden in the singularity.  It is interesting to consider the
relation between these two kinds of object.

Consider the action for the metric, an NS-NS field strength, and an R-R field
strength, schematically
\be
S = \int d^Dx\, \left( e^{-2\phi} R -  e^{-2\phi} H^2 -  G^2
\right) \ .  \label{gnract}
\ee
For a black hole (or more generally $p$-brane) carrying NS-NS charge, the
solution is determined by the balance between the first two terms.  Since the
dilaton appears in the same way in both terms, the classical solution is
independent of the dilaton.  That is, its size is independent of the string
coupling and its mass scales as the action,
$e^{-2\phi} = g^{-2}$. For a black $p$-brane carrying R-R charge, the solution
is determined by a balance between the first and third terms.  Its mass is
then the mean,
$e^{-\phi} = g^{-1}$, smaller at weak coupling than that of the NS-NS object. 
Similarly, one can see by a scaling argument that the size goes to zero with
the string coupling.  For couplings less than one, the classical solution is
smaller than the string scale.  At this point one's prejudice from
perturbative string theory is that geometry, or at least the low energy
effective action we are using, makes no sense and the black solution is not
relevant.
Instead, as with the earlier example of instantons, for solitons smaller than
the string scale D-branes provide the correct effective description.

One rather prosaic scenario for what all this means is that string theory
really can be defined as some closed string field theory, and the D-branes are
just solitons of this theory.  The low energy action~(\ref{gnract}) is not
valid for these solitons, but the full closed string action would correctly
describe the D-branes.  My own prejudice\,\cite{lesh} is that string field
theory is not likely to be a useful description beyond perturbation theory.
String duality has reinforced this, in that each string theory
appears to be an effective theory with a very specific range of validity.

It is interesting to consider now the case of $Q$ D-branes.  Each boundary on
the string world-sheet then brings in a factor of $gQ$, so the condition for
validity of the D-brane approximation is now $gQ < 1$.  At the same time, one
finds that the Schwarzchild radius in string units is also $gQ$, so the
low energy geometric description should be valid for $gQ > 1$.

Using a standard strategy in string duality we can continue in $g$ and the
number of supersymmetric (BPS) states must be invariant.  Strominger and Vafa
applied this to a collection of D-branes chosen in such a way that the
resulting black hole would have a nonzero area horizon, finding that the
density of states in the D-brane system corresponded precisely to that given
by the black hole entropy.\cite{sv}  Because of the close analogy between
black hole mechanics and thermodynamics, there have been many attempts to
give a statistical mechanical interpretation to the black hole entropy, but
this is the first time that it has been done in a controlled way.\footnote
{Larsen and Wilczek\,\cite{larwil} had applied the same strategy to an NS black
hole, but here the system remains a black hole even at weak coupling and so
there is no explicit understanding of the space of states.}

This has of course led to a great surge of interest.  Unfortunately this
is beyond the scope of these lectures, but I should at least discuss in brief
the implications for the black hole information paradox.  At a superficial
level there are none: the success of the state counting might be expected
regardless.  We have two descriptions with different ranges of validity, and
while the D-brane system has an explicit quantum mechanical description, one
could certainly imagine that at
$gQ \sim 1$ the system makes a transition from being described to high accuracy
by ordinary quantum mechanics to one with information loss.

Nevertheless, it
is worth pushing things further and asking whether we can say more about black
hole quantum mechanics from D-branes.  Indeed, there are some successes that
seem to go beyond what supersymmetry alone would require.  These
include\,\footnote
{I apologize in advance for identifying only a small number of papers out of
a large literature, but these seem to have been highlights at least in a
certain direction of development.} counting of states of some nonextremal
black holes\,\cite{nonex} and comparisons of the decay rates of black holes and
D-branes\,\cite{decay} even at a fairly high level of detail.\,\cite{grey} 
Yet, even while there is some indication that black holes might be described by
ordinary quantum mechanics, it is important to remember that there is a paradox
here.  Twenty years of attempts to show that information is preserved in black
hole evaporation have only served to point up that if it is so it will imply
a significant change in our way of thinking about spacetime.  Enthusiasts of
D-brane--black hole quantum mechanics should keep their eyes open for the
important lesson.

\subsection{Conclusion}

The Theory of Everything---string theory, M-theory, whatever it should
eventually be called---is now understood as in figure~13.
\begin{figure}
\begin{center}
\leavevmode
\epsfbox{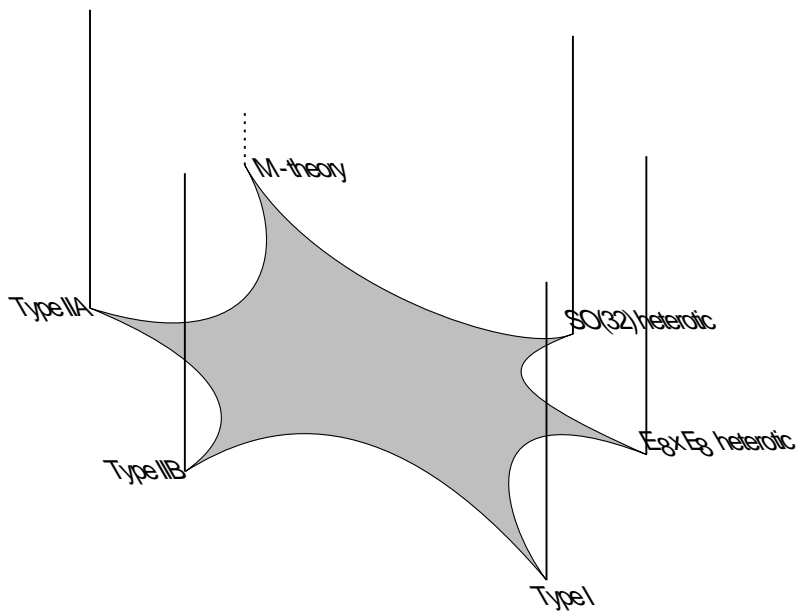}
\end{center}
\caption[]{All string theories, and M-theory, as limits of one theory. 
Energy, or more generally  distance from the BPS limit, increases vertically.}
\end{figure}
The various string theories are asymptotic expansions around the 
cusps.  In addition, supergravity provides an effective description valid in
the whole picture but only at low energy, and one can follow the
BPS states from one limit to another.

This is far from a nonperturbative definition of the theory.  I have no doubt
that such a definition is possible, and that (as with the renormalization
group in quantum field theory) it is a necessary step towards understanding
the dynamics of the theory. Where do D-branes fit in?  At a superficial
level, they merely provide, in the Type I and Type II limits, a more precise
description of the R-R solitons in the spectrum.  Yet it may be that in fact
they are closer than strings to being the fundamental degrees of freedom
needed to define the whole picture.  The fact that they probe shorter
distance scales\,\cite{short} and perhaps the unexpected successes of 
D-brane--black hole quantum mechanics described above suggest that this may
be so.  Thus, the D-brane Hamiltonian, or some structure abstracted from
it, may have a much greater range of validity than the perturbative
string theory in which it was first described.\cite{bfss}

\section*{Acknowledgments}
I would like to thank Shyamoli Chaudhuri and Clifford Johnson for their
collaboration on the earlier version of these notes, and Shyamoli for her
careful reading of the manuscript.  I would also like to thank Martin
Rocek and the other members of the Stony Brook D-Brane Reading Group for
their detailed comments and suggestions.  This work is supported by NSF grants
PHY91-16964 and PHY94-07194.


\section*{References}
\begin{thebibliography}{99}

\bibitem{dlp} 
J.~Dai, R.~G.~Leigh and J.~Polchinski, Mod.~Phys.~Lett.
{\bf A4} (1989) 2073.

\bibitem{joeone}
J.~Polchinski, Phys.~Rev. Lett.~{\bf 75} (1995) 4724.

\bibitem{dnotes}  
J. Polchinski, S. Chaudhuri, and C. Johnson, {\it Notes on
D-Branes,} preprint NSF-ITP-96-003, hep-th/9602052.

\bibitem{early}
A. Chodos and C. B. Thorn, Nucl. Phys. {\bf B72} (1974) 509;\hfil\break
W. Siegel, Nucl. Phys. {\bf B109} (1976) 244;\hfil\break
S. M. Roy and V. Singh, Pramana {\bf 26} (1986) L85; Phys. Rev. {\bf D35}
(1987) 1939.

\bibitem{latera}
J. A. Harvey and J. A. Minahan, Phys. Lett. {\bf B188} (1987) 44.

\bibitem{laterc}
N. Ishibashi and T. Onogi, Nucl. Phys. {\bf B318} (1989) 239;\hfil\break
G. Pradisi and A. Sagnotti, Phys. Lett. {\bf B216} (1989) 59;\hfil\break
A. Sagnotti, Phys. Rept. {\bf 184} (1989) 167;\hfil\break 
P. Horava, Nucl. Phys. {\bf B327} (1989) 461.

\bibitem{hdual}
P. Horava, Phys. Lett. {\bf B231} (1989) 251.

\bibitem{gdual}
M. B. Green, Phys. Lett. {\bf B266} (1991) 325. 

\bibitem{offshell}
J. H. Schwarz, Nucl. Phys. {\bf B65} (1973), 131;\hfil\break  
E. F. Corrigan and D. B. Fairlie, Nucl. Phys. {\bf B91} (1975)
527;\hfil\break 
M. B. Green, Nucl. Phys. {\bf B103} (1976) 333;\hfil\break 
M. B. Green and J. A. Shapiro, Phys. Lett. {\bf 64B} (1976) 454;\hfil\break 
A. Cohen, G. Moore, P. Nelson, and J. Polchinski, Nucl. Phys. {\bf B267}, 143
(1986); {\bf B281}, 127 (1987).

\bibitem{parton}
M. B. Green, Phys. Lett. {\bf B69} (1977) 89; {\bf
B201} (1988) 42; {\bf B282} (1992) 380.

\bibitem{joecomb}
J.~Polchinski, Phys.~Rev.~{\bf D50} (1994) 6041.

\bibitem{shenk1}
S. H. Shenker, in {\it Cargese 1990, Proceedings:  Random Surfaces and Quantum
Gravity} (1990) 191.

\bibitem{gdinst}
M. B. Green, Phys. Lett. {\bf B329} (1994) 435.

\bibitem{chan}
J. Paton and Chan Hong-Mo, Nucl. Phys. {\bf B10} (1969)
519.

\bibitem{sms}
J. H. Schwarz, in {\it Florence 1982, Proceedings, Lattice Gauge Theory,
Supersymmetry and Grand Unification,} 233; Phys. Rept. {\bf 89} (1982)
223;\hfil\break
N. Marcus and A. Sagnotti, Phys. Lett. {\bf 119B} (1982) 97.

\bibitem{opentwist}
A. Sagnotti, {\it
Non-Perturbative Quantum Field Theory,}
eds. G. Mack et. al. (Pergamon Press, 1988) 521;\hfil\break
V. Periwal, unpublished;\hfil\break
J. Govaerts, Phys. Lett. {\bf B220} (1989) 77.

\bibitem{mirror}
A. Strominger , S.-T. Yau, and E. Zaslow, {\it Mirror Symmetry is
$T$-Duality,} hep-th/9606040;\hfil\break
H. Ooguri, Y. Oz, and Z. Yin, Nucl. Phys. {\bf B477} (1996) 407, 
hep-th/9606112.

\bibitem{tdual}
K. Kikkawa and M. Yamanaka, Phys. Lett. {\bf B149} (1984) 357;\hfil\break 
N. Sakai and I. Senda, Prog. Theor. Phys. {\bf 75} (1986) 692.

\bibitem{nairet}
V.P. Nair, A. Shapere, A. Strominger, and F. Wilczek,
Nucl. Phys. {\bf B287} (1987) 402.

\bibitem{dhs}
M. Dine, P. Huet, and N. Seiberg, Nucl. Phys. {\bf
B322} (1989) 301.

\bibitem{gvb}
P. Ginsparg and C. Vafa, Nucl. Phys. {\bf B289} (1987) 414;\hfil\break 
T. H. Buscher, Phys. Lett. {\bf B194B} (1987) 59; {\bf B201} (1988) 466. 

\bibitem{witbound}
E. Witten, Nucl. Phys. {\bf B460} (1996) 335, hep-th/9510135.

\bibitem{leigh}
R.~G.~Leigh, Mod.~Phys.~Lett. {\bf A4} (1989) 2767.

\bibitem{bachas}
C. Bachas, Phys. Lett. {\bf B374} (1996) 37, hep-th/9511043.

\bibitem{tdbi}
E. Bergshoeff, M. de Roo, M. B. Green, G. Papadopoulos, and P.K.
Townsend, Nucl. Phys. {\bf B470} (1996) 113, hep-th/9601150;\hfil\break
E. Alvarez, J. L. F. Barbon, and J. Borlaf, {\it $T$-Duality for Open
Strings,} hep-th/9603089;\hfil\break
E. Bergshoeff and M. De Roo, Phys. Lett. {\bf B380} (1996) 265,
hep-th/9603123.

\bibitem{frad} 
E. S. Fradkin and A. A. Tseytlin, Phys.
Lett. {\bf B163} (1985) 123.

\bibitem{tdtp}
M. B. Green, C. M. Hull, and P. K. Townsend,
Phys. Lett. {\bf B382} (1996) 65, hep-th/9604119.

\bibitem{shanta}
S.P. de Alwis, {\it A Note on Brane Tension and M Theory,}
hep-th/9607011.

\bibitem{chans}
C. Lovelace, Phys. Lett. {\bf B34} (1971) 500;\hfil\break
L. Clavelli and J. Shapiro, Nucl. Phys. {\bf B57} (1973) 490;\hfil\break
M. Ademollo, R. D' Auria, F. Gliozzi, E. Napolitano, S. Sciuto, and P. di
Vecchia, Nucl. Phys. {\bf B94} (1975) 221;\hfil\break
C. G. Callan, C. Lovelace, C. R. Nappi, and S. A. Yost,
Nucl. Phys. {\bf B293} (1987) 83.

\bibitem{rrex}
J. Polchinski and Y. Cai, Nucl. Phys. {\bf B296} (1988) 91;\hfil\break 
C. G. Callan, C. Lovelace, C. R. Nappi and S.A. Yost,
Nucl. Phys. {\bf B308} (1988) 221.

\bibitem{cw}
S. Coleman and E. Weinberg, Phys. Rev. {\bf D7}
(1973) 1888;\hfil\break 
J. Polchinski, Comm. Math. Phys. {\bf 104}
(1986) 37.

\bibitem{occoup}
J. A. Shapiro and C. B. Thorn, Phys.
Rev. {\bf D36} (1987) 432;\hfil\break J. Dai and J. Polchinski, Phys. Lett.
{\bf B220} (1989) 387.


\bibitem{sobig}
M. Douglas and B. Grinstein, Phys. Lett. {\bf B183} (1987) 552; (E) {\bf 187}
(1987) 442;\hfil\break 
S. Weinberg, Phys. Lett. {\bf B187} (1987) 278;\hfil\break
N. Marcus and A. Sagnotti, Phys. Lett. {\bf B188} (1987) 58.

\bibitem{fms}
D. Friedan, E. Martinec, and S. Shenker, Nucl. Phys. {\bf B271}
(1986) 93.

\bibitem{blackp}
G. T. Horowitz and A. Strominger,
Nucl. Phys. {\bf B360} (1991) 197.

\bibitem{hullt}
C. M. Hull and P.K. Townsend,
Nucl. Phys. {\bf B438} (1995) 109.

\bibitem{town}
P. K. Townsend, Phys. Lett. {\bf B350} (1995) 184, 
hep-th/9501068.

\bibitem{wit}
E. Witten, Nucl. Phys. {\bf B443} (1995) 85.

\bibitem{coni}
A. Strominger, Nucl. Phys. {\bf B451} (1995) 96.

\bibitem{romans}
L. J. Romans, Phys. Lett. {\bf B169} (1986) 374.

\bibitem{joeandy}
J. Polchinski and A. Strominger, {\it New Vacua for Type II String Theory,}
hep-th/9510227.

\bibitem{rract}
M. Li, Nucl. Phys. {\bf B460} (1996) 351, hep-th/9510161.

\bibitem{douginst}
M. R. Douglas, {\it Branes within Branes,} hep-th/9512077.

\bibitem{bsv}
M. Bershadsky, C. Vafa, and V. Sadov, Nucl. Phys. {\bf B463}
(1996) 420, hep-th/9511222.

\bibitem{ghs}
M. B. Green, J. A. Harvey, and
G. Moore, {\it I-Brane Inflow and Anomalous Couplings on
D-Branes,} hep-th/9605033.

\bibitem{bpspict}
M. Bianchi, G. Pradisi, and A. Sagnotti, Nucl. Phys. {\bf B376} (1992)
365.

\bibitem{tomlen}
T. Banks and L. Susskind, {\it Brane - Anti-Brane Forces,}
hep-th/9511194.

\bibitem{fsuss}
W. Fischler and L.
Susskind, Phys. Lett. {\bf B171} (1986) 383; {\bf 173} (1986) 262.

\bibitem{gsdiv}
M. B. Green and J. H. Schwarz, Phys. Lett.
{\bf B149} (1984) 117; {\bf B151} (1985) 21.

\bibitem{weinberg}
S. Weinberg, {\it The Quantum Theory of Fields.  Vol. 1: Foundations,}
(Cambridge U. P., Cambridge UK, 1995). 

\bibitem{dirac}
R. I. Nepomechie, Phys. Rev. {\bf D31}
(1985) 1921;\hfil\break 
C. Teitelboim, Phys. Lett. {\bf B167} (1986) 63, 69.

\bibitem{bdl}
M. Berkooz, M. R. Douglas, and R. G. Leigh, {\it Branes Intersecting at
Angles,} hep-th/9606139.

\bibitem{witinst}
E. Witten, Nucl. Phys. {\bf B460} (1996) 541, hep-th/9511030.

\bibitem{GP}
E. G. Gimon and Joseph Polchinski, Phys. Rev. {\bf D54}
(1996) 1667, hep-th/9601038.

\bibitem{schwarz}
J. H. Schwarz, Phys. Lett.
{\bf B360} (1995) 13; (E) {\bf B364} (1995) 252.

\bibitem{towndf}
P .K. Townsend, Phys. Lett. {\bf B373} (1996) 68, hep-th/9512062.

\bibitem{senbound}
A. Sen, Phys. Rev. {\bf D54} (1996) 2964.

\bibitem{senbound2}
A. Sen, Phys. Rev. {\bf D53} (1996) 2874.

\bibitem{vafa1}
C. Vafa, Nucl. Phys. {\bf B463} (1996) 415, hep-th/9511088.

\bibitem{horpriv} 
G. Horowitz, private communication.

\bibitem{08}
G. Papadopoulos and P. K. Townsend, {\it Kaluza-Klein on the
Brane,} hep-th/9609095;\hfil\break
U. H. Danielsson and G. Ferretti, {\it The Heterotic Life of the D-particle,}
hep-th/9610082.

\bibitem{hullscale}
C. M. Hull, Nucl. Phys. {\bf B468} (1996) 113, hep-th/9512181.

\bibitem{hetsol}
J. A. Harvey and A. Strominger, Nucl. Phys. {\bf B449} (1995) 535;\hfil\break
A. Sen, Nucl. Phys. {\bf B450} (1995) 103.
 
\bibitem{polwit}
J. Polchinski and E. Witten, Nucl. Phys. {\bf B460} (1996) 525, hep-th/9510169.

\bibitem{blackone}
A. Dabholkar,  Phys. Lett. {\bf B357} (1995) 307;\hfil\break 
C. M. Hull, Phys. Lett. {\bf B357} (1995) 545.

\bibitem{supmem}
E. Bergshoeff, M. J. Duff, C. N. Pope, and E.
Sezgin, Phys. Lett. {\bf B224} (1989) 71.

\bibitem{horwit}
P. Horava and E. Witten, Nucl. Phys. {\bf B460} (1996) 506, hep-th/9510209.

\bibitem{duff11}
M. J. Duff and J. X. Lu, Nucl. Phys. {\bf B390} (1993) 276, hep-th/9207060;\\
C. Schmidhuber, Nucl .Phys. {\bf B467} (1996) 146, hep-th/9601003;\\
S. P. de Alwis and K. Sato, Phys. Rev. {\bf D53} (1996) 7187,
hep-th/9601167;\\
A. A. Tseytlin, Nucl. Phys. {\bf B469} (1996) 51, hep-th/9602064.

\bibitem{narain}
K. S. Narain, Phys. Lett. {\bf 169B} (1986) 41;\\
P. Ginsparg, Phys. Rev. {\bf D35} (1987) 648. 

\bibitem{julia}
B. Julia, in {\it Supergravity and Superspace,} ed. by S. W. Hawking and M.
Rocek (Cambridge U. P., Cambridge UK, 1981).

\bibitem{chs}
C. G. Callan, J. A. Harvey, and A. Strominger, Nucl. Phys. {\bf
B359} (1991) 611; {\bf B367} (1991) 60.

\bibitem{vafwit}
C. Vafa and E. Witten, Nucl. Phys. {\bf B431} (1994) 3,
hep-th/9408074.

\bibitem{vafa2}
C. Vafa, Nucl. Phys. {\bf B463} (1996) 435, hep-th/9512078.

\bibitem{andyopen}
A. Strominger, Phys. Lett. {\bf B383} (1996) 44, hep-th/9512059. 

\bibitem{dougprobe}
M. Douglas, {\it Gauge Fields and D-Branes,} hep-th/9604198;\\
M. R. Douglas and G. Moore, {\it D-Branes, Quivers, and
ALE Instantons,} hep-th/9603167.

\bibitem{witadmh}
E. Witten, J. Geom. Phys. {\bf 15} (1995) 215, hep-th/9410052.
 
\bibitem{admh}
M. F. Atiyah, V. G. Drinfeld, N. J. Hitchin, and Y. I. Manin, Phys. Lett.
{\bf A65} (1978) 185.

\bibitem{shenk2}
S. H. Shenker, {\it Another Length Scale in String Theory?} preprint RU-95-53,
hep-th/9509132 and seminar at ITP, Jan. 1996.

\bibitem{short}
D. Kabat and P. Pouliot, {\it A Comment on Zero-Brane Quantum
Mechanics,} hep-th/9603127;\\
U. H. Danielsson, G. Ferretti, and B. Sundborg, 
{\it D-Particle Dynamics and Bound States,} hep-th/9603081;\\
G. Lifschytz, {\it Comparing D-branes to Black-branes,}
hep-th/9604156;\\
M. R. Douglas, D. Kabat, P. Pouliot and S. H. Shenker,
{\it D-branes and Short Distances in String Theory,} hep-th/9608024.

\bibitem{tensors}
J. Polchinski, {\it Tensors from $K3$ Orientifolds,} hep-th/9606165.

\bibitem{ale} 
T. Eguchi and A. J. Hanson, Ann. Phys. {\bf 120} (1979) 82;\\
G. W. Gibbons and S. W. Hawking, Comm. Math. Phys. {\bf 66} (1979) 291.

\bibitem{hyper} 
A. Hitchin, A. Karlhede, U. Lindstrom, and M. Ro\v cek,
Comm. Math. Phys. {\bf 108} (1987) 535;\\
P. Kronheimer, J. Diff. Geom. {\bf 28} (1989) 665; {\bf 29} (1989) 685.

\bibitem{cvjmyers}
C. V. Johnson and R. C. Myers, {\it Aspects of Type IIB Theory on ALE Spaces,}
hep-th/9610140.

\bibitem{lesh}
J. Polchinski, ``What is String Theory?,'' 
Proceedings of the 1994 Les Houches Summer
School, edited by F. David, P. Ginsparg and J. Zinn-Justin
(Elsevier, Amsterdam, 1996) p. 287,
hep-th/9411028.

\bibitem{sv}
A. Strominger and C. Vafa, {\it Microscopic Origin of
the Bekenstein-Hawking Entropy,} hep-th/9601029.

\bibitem{larwil}
F. Larsen and F. Wilczek, Phys. Lett. {\bf B375} (1996) 37.

\bibitem{nonex}
C. G. Callan, Jr. and J. M. Maldacena, Nucl. Phys. {\bf B472}
(1996) 591, hep-th/9602043;\\
G. T. Horowitz and A. Strominger, Phys. Rev. Lett. {\bf 77}
(1996) 2368, hep-th/9602051;\\
G. T. Horowitz, J. M. Maldacena, and
Andrew Strominger, Phys. Lett. {\bf B383} (1996) 151, hep-th/9603109;\\
J. M. Maldacena, {\it Black Holes in String Theory,} Ph.~D. Thesis,
hep-th/9607235.

\bibitem{decay}
A. Dhar, G. Mandal and S. R. Wadia, Phys. Lett. {\bf B388} (1996) 51,
hep-th/9605234;\\
S. R. Das and Samir D. Mathur, {\it Comparing Decay Rates for Black Holes and
D-Branes,} hep-th/9606185;\\
S. R. Das and S. D. Mathur, {\it Interactions Involving D-Branes,}
hep-th/9607149. 

\bibitem{grey}
J. Maldacena and A. Strominger, {\it Black Hole Grey Body Factors and D-Brane
Spectroscopy,} hep-th/9609026;\\
C. G. Callan, Jr., S. S. Gubser, I. R. Klebanov, and A. A. Tseytlin,
{\it Absorption of Fixed Scalars and the D-Brane Approach to Black Holes,}
hep-th/9610172.

\bibitem{bfss}
T. Banks, W. Fischler, S. H. Shenker, and L. Susskind, {\it M Theory as a Matrix
Model: A Conjecture,} hep-th/9610043.

\end{thebibliography}
\end{document}